# Possible Signals of  new phenomena  in hadronic interactions at $dn/d\eta = 5.5 \pm 1.2$

R.Campanini

*Abstract*—The average transverse momentum dependence on multiplicity shows in many experiments at center of mass energies ranging from 22 to 7000 GeV a slope change at a charged particle rapidity density constant within systematic uncertainties. We find correlated signals which together with the slope change may indicate a transition to a new mechanism of particles production.

## I.  INTRODUCTION.

THE average transverse momentum <$p_t$> has been studied by many experiments as function of the charged particle rapidity density $\rho = dn/dy$. In the last 25 years many experiments have shown that there is an increase in <$p_t$> with ρ in the central rapidity region. A number of  possible explanations for this increase have been proposed. Among these are geometrical models (1),"mini-jets" production (2), and thermodynamical models(3). In (4) we pointed out that $pp$ and $p\bar{p}$ data showed a kind of jump at ρ=6 and that it had to be investigated as a possible phase transition signal. In this work we review many experimental results in hadron-hadron collisions, including recent LHC data. We show that in different experiments, at various energies ranging from $\sqrt{s}$ = 22 GeV up to 7 TeV, the <$p_t$> vs ρ dependences show a slope change in a well defined ρ region. The slope change at very different energies appears at ρ value ρ* which is stable within  systematic uncertainties. We observed that many features of the events depend on R=ρ/ρ*. We put forward the hypothesis that the slope change in the <$p_t$> vs ρ together with correlated signals may indicate a  transition to a new particle production mechanism. The average of  p* on all the considered experiments  is 5.5 with standard deviation of 0.6. In (5), Alexopoulos et al. assumed that the system produced in $p\bar{p}$ at $\sqrt{s}$=1800 GeV for ρ >6.75 was above the deconfinement transition to explain their experimental results. From our finding it may be that at  ρ = 5.5 ± 1.2  there is a critical point independent of $\sqrt{s}$.

## II.  INCREASE IN <$P_t$> WITH THE CHARGED PARTICLE DENSITY IN RAPIDITY.

To our knowledge the increase of <$p_t$> with the multiplicity in elementary hadronic interactions has been observed at the lowest energy by the NA22 collaboration (6), at center of mass energy $\sqrt{s}$ = 22 GeV in k$p$, π$p$ and  $pp$ interactions. The study has been done in different rapidity intervals, with different cuts on the lowest used $p_t$ ($p_{t\,min}$). In fig. 1 we show  the results for -2.0 < y < 2.0. The <$p_t$> increases when $p_{t\,min}$> 0.3 GeV/c or $p_{t\,min}$> 0.4 GeV/c cuts are applied. The charged multiplicity n is the number of particles for which the <$p_t$> has been calculated. The <$p_t$> increase stops at multiplicity n = 8. To correct multiplicity for  $p_{t\,min}$  cut, we consider an exponential law  for  the  invariant  $p_t$  distribution $dn/dp_t{}^2 = A\,e^{-p_t/\overline{p_t}}$ with  $\overline{p_t}$  = 0.370  GeV/c  and we extrapolate to  $p_t$ =0. The  corrected charged particle rapidity density where the < $p_t$ > increase stops is ρ = 5.5. Uncertainty sources are the multiplicity bin size  and the uncertainty on correction for the $p_{t\,min}$  cut .We estimated a total uncertainty of ± 0.5 with a level of confidence  of approximately 95%.At higher rapidity densities, statistics are very poor and no significant conclusions for n > 8 can be drawn.

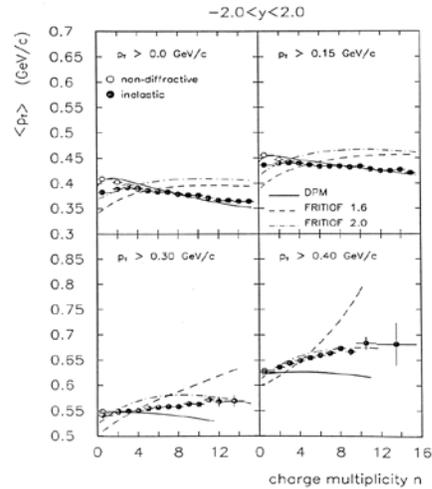

Fig. 1  <$p_t$> vs charged multiplicity n  at $\sqrt{s}$=22 GeV  (6).Lines are models predictions in the NA22  paper.

At Intersecting Storage Ring ISR at CERN the <$p_t$> vs ρ has been studied at $\sqrt{s}$ of 31,44, and 63 GeV, in $pp$, $p\bar{p}$, $\alpha\alpha$, and

Author is with the Dipartimento di Fisica dell'Università di Bologna and I.N.F.N. Sezione di Bologna,Italy ( e-mail: renato.campanini@ unibo.it).

$p\alpha$ interactions. In (7) (ABCDHW coll.) the analysis was based on 880 000, 460 000 and 1 400 000 "minimum bias" events at $\sqrt{s}$ of 62, 44 and 31 GeV respectively and 350 000 "electron trigger" events at $\sqrt{s} = 62$ GeV. The "electron-trigger" selected events with an electron candidate of positive or negative charge produced at a polar angle of 90 degrees. The multiplicity of "electron-trigger" events was 2.5 times the minimum bias multiplicity at the same energy. Results have been presented both for rapidity interval -1.5 < y < 1.5 and -0.5 < y < 0.5, with $0.150 \le p_t \le 2.5$ GeV/c. For the "electron-trigger" particle <$p_t$> a $0.4 < p_t < 1.0$ GeV/c cut has been applied to reduce hadron contamination. The $\rho$ densities have not been corrected for apparatus acceptance, which has been estimated to be $\cong$ 80%. The <$p_t$> has been estimated from the arithmetic mean. Acceptance corrections have been computed via Monte Carlo simulation over limited phase-space regions covered by the apparatus and check has been done that the <$p_t$> dependence on $\rho$ was essentially unchanged using acceptance corrected or not corrected values. Results are shown from fig.2 to fig.5. The mean transverse momentum increases with $\rho$ at all energies. The rise of <$p_t$> becomes steeper when $\sqrt{s}$ increases from 31 to 62 GeV. Taking the lower boundary of the $p_t$ interval to be $p_t = 0.300$ GeV/c further enhances the rise of <$p_t$> with $\rho$ (not shown). In the "electron-trigger" events, an increase of <$p_t$> with increasing $\rho$ is observed for both the trigger and the non trigger particles. In all data samples the rise is approximately linear up to $\rho$ around 4 for $|y| < 1.5$ and 6 for $|y| < 0.5$, where a slope change is visible. An average correction factor of $\cong$ 1.25 for acceptance losses shifts the $\rho = 4.0$ to $\rho \cong 5$. Uncertainty sources are the approximate correction for acceptance and the $\rho$ bin size. We estimate the total uncertainty to be $\pm$ 0.5 at 95% confidence level. A Monte Carlo study has been done (8) in order to see if the increases were due to phase space effects, which was not the case.

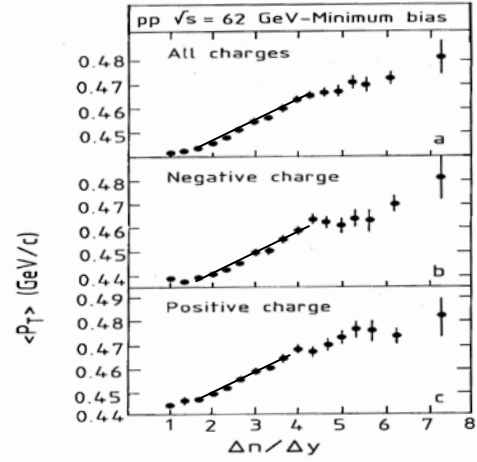

Fig. 3 <$p_t$> vs dn/dy, $|y| < 1.5$ , $0.150 < p_t < 2.5$ GeV/c in pp minimum bias $\sqrt{s} = 62$ GeV. Data are not corrected for acceptance (7) and are plotted together with a pre-selected "eye-guide" line.

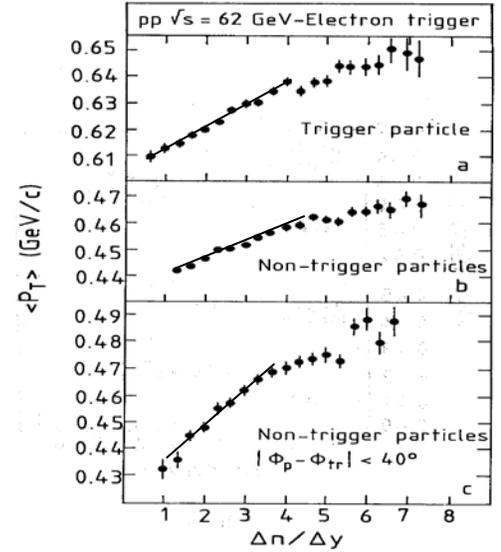

Fig. 4 <$p_t$> vs dn/dy, "electron-trigger" data $\sqrt{s} = 62$ GeV, ABCDHW coll.(7). Data are not corrected for acceptance and are plotted together with a pre-selected "eye-guide" line.

(a)  The average transverse momentum of the triggering particle with $0.4 < p_t < 1$ GeV/c as a function of $\rho = dn/dy$;

(b)  The average transverse momentum of all non-trigger particles having $0.15 \le p_t \le 2.5$ GeV/c as a function of $\rho$;

(c)  The average transverse momentum for non-trigger particles, with $0.15 \le p_t \le 2.5$ GeV/c, lying in a $\Phi$ interval of $\pm$ 40 degrees around the triggered track.

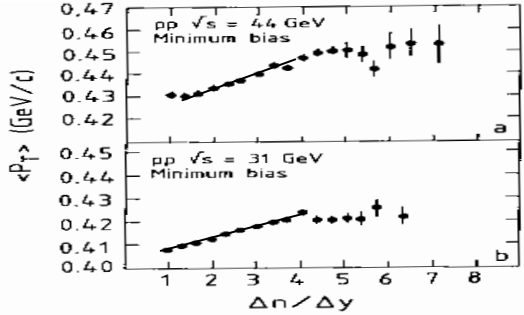

Fig. 2. < $p_t$ > vs $\frac{dn}{dy}$ for $|y| < 1.5$, $0.150 < p_t < 2.5$ GeV/c , pp minimum bias, $\sqrt{s} = 44$, 31 GeV. Data are not corrected for acceptance (7) and are plotted together with a pre-selected "eye-guide" line.

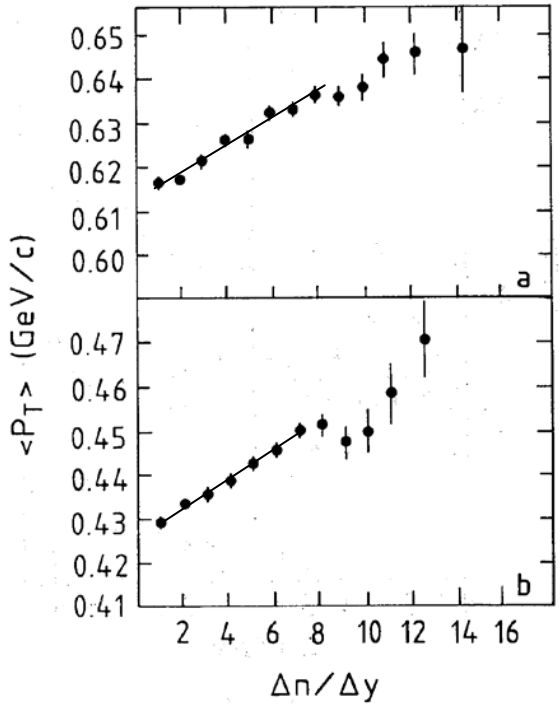

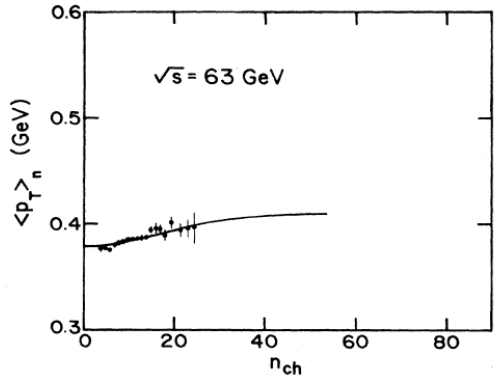

Fig. 5 (a) Electron trigger data. The average transverse momentum of the triggering track as a function of the particle density computed for y interval ± 0.5 around the triggering track rapidity. (b) Minimum bias data at $\sqrt{s}$ =62 GeV. The average transverse momentum for tracks having $0.15 \leq p_t \leq 2.5$ GeV/c and |y|<0.5 as a function of the particle density in the same y interval (7). Data are not corrected for acceptance and are plotted together with a pre-selected "eye-guide" line.

In fig.6 one can see results of minijets model prediction(9) for the 63 GeV data.We will return later on minijets model predictions.

Fig. 6  Results of minijets model calculation for the ABCDHW data at $\sqrt{s}$=63 GeV.From (9)

It is interesting to know how the $p_t$ spectra vary with multiplicity. In fig.7 from (10), the ratio between the $p_t$ spectrum at a given multiplicity and the inclusive spectrum is plotted. For low multiplicity events, a bump is seen for $p_t$ around 0.4 GeV/c, whereas at large multiplicity a dip is seen in the same $p_t$ range. These changes indicate a widening of the dn/d$p_t$ distributions for increasing multiplicity n. The distribution becomes flatter at $p_t$ > 0.4 GeV/c, but also rises at low $p_t$ ($p_t$ <0.25): this kind of change explains, at least at ISR energies, the fact that experimental cuts in the $p_t$ spectrum strongly influence the multiplicity dependence of <$p_t$>. It is very interesting that for multiplicities $n_c$> 18, which correspond in the experiment to ρ > 5.6 there is an increase of production of low $p_t$ particles ($p_t$ <0.25). The <$p_t$> increase with multiplicity is due to the region for $p_t$> 0.7 GeV/c.

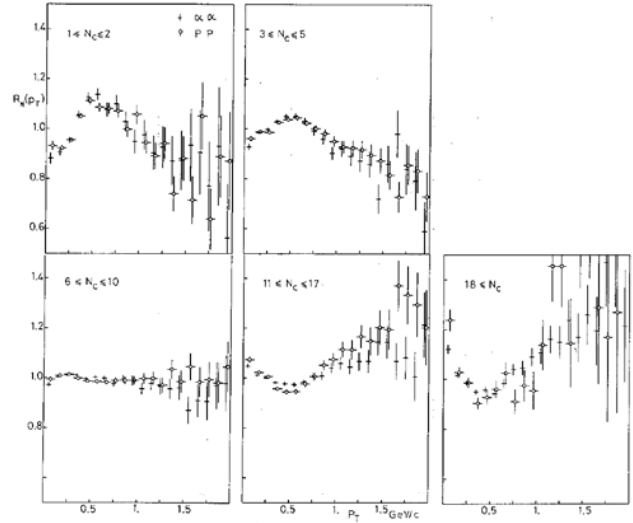

Fig. 7. Normalized ratio of the transverse momentum distributions for different multiplicity bins for α α ($\sqrt{s_{NN}}$ = 31.2 GeV) and $pp$ ($\sqrt{s}$ =63 GeV) (10)

At SPS collider the <$p_t$> vs ρ dependence has been measured by the UA1 collaboration (11). In fig. 8, 9, 10 results are shown respectively at $\sqrt{s}$ of 200, 540 and 900 GeV, together with the prediction of a minijet model (9). The <$p_t$> has been calculated with $p_{t\,min}$ cut > 0.150 GeV/c with extrapolation to $p_t$ = 0 with a power law. The particles pseudorapidity is in the interval -2.5 < η < 2.5. The pseudorapidity densities are acceptance corrected; the multiplicity systematic error is ± 12%. A clear slope change is present in the <$p_t$> vs $n_{ch}$ multiplicity at all energies, at $n_{ch}$ =28 which corresponds to ρ ≡dn/dη = 5.6. From the multiplicity systematic error we get an uncertainty of ± 0.6 on the ρ value.Multiplicity bin size is another source of uncertainty of ± 0.4. We can estimate a total uncertainty at approximately 95% confidence level of ± 0.9. As we can see, at all energies the minijets model describes well the rise, but it does not reproduce the slope change. Also the HIJING model (12) fails to describe the slope change, as seen in fig.11. The minijets model of (2) at 540 GeV gives a general good description of the increase but again does not reproduce the slope change (see fig.12). In fig.13 UA1 results at 630 GeV (13) from 2.36 million minimum bias $p\bar{p}$ events are shown. The <$p_t$> has been calculated in the range 0.180 < $p_t$ <25GeV/c , with extrapolation to 0 with the power law formula

$$E\frac{d^3\sigma}{dp^3} = \frac{Ap_t^{0^n}}{(p_t + p_t^0)^n}$$

with 2 different values of $p_t^0$. Data are acceptance corrected. The multiplicity systematic error is ±12%. There is a slope change at $dn/d\eta = 7.6$. The uncertainty on this value is due to multiplicity systematic error ($\pm$ 12%) and bin size($\pm$ 0.5).We can estimate a total uncertainty of $\pm$ 1.0, at approximately 95% confidence level. Authors point out that the $p_t$ spectra calculated at different multiplicities become nearly parallel for $p_t > 2.2$ GeV/c. This suggests that the rise of $<p_t>$ with multiplicity occurs because there are relative fewer particles at low $p_t$ in high multiplicity events. The effect may be then associated with changing production of lower than 2.2 GeV/c $p_t$ particles.

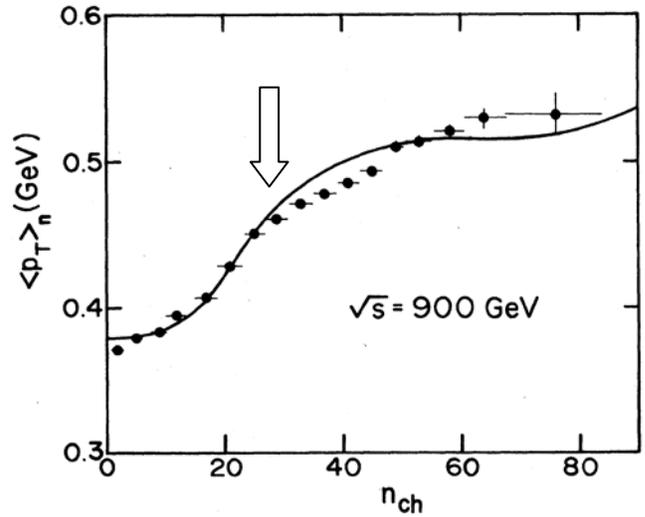

Fig. 10 As in fig.8 , $\sqrt{s}$ = 900 GeV (9). Arrow points to slope change.

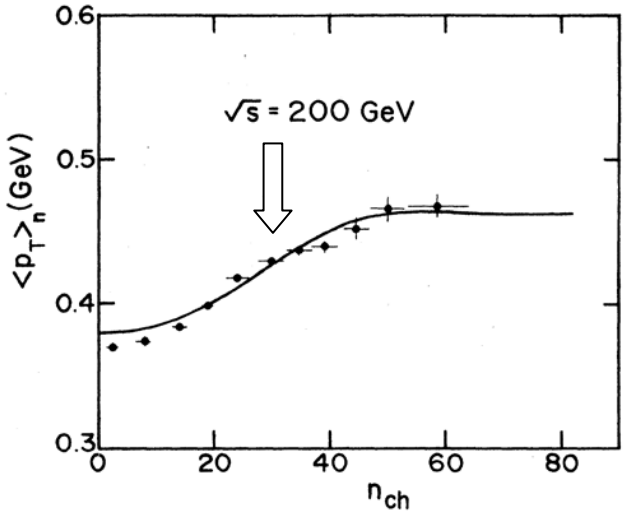

Fig. 8 $<p_t>$ ns charge multiplicity $n_{ch}$, $p\bar{p}$ minimum bias, $\sqrt{s}$ = 200 GeV, UA1 coll., with minijet model prediction (9). Arrow points to slope change.

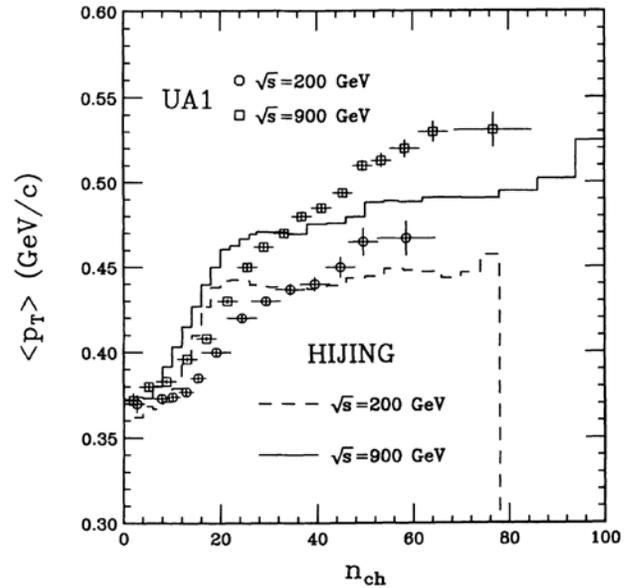

Fig. 11 HIJING (1992 version) compared to UA1 data at $\sqrt{s}$ = 200 and 900 GeV (12).

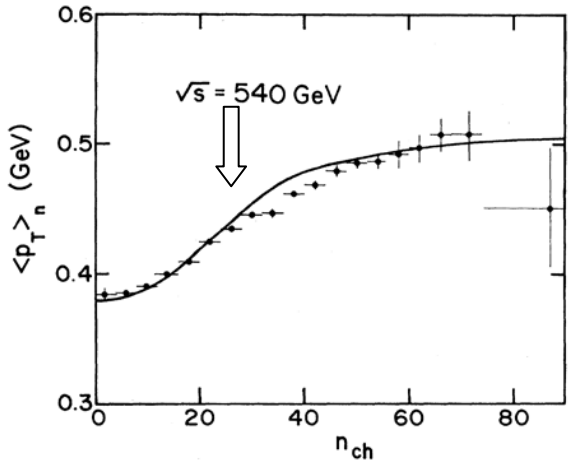

Fig. 9 As in fig 8 for $\sqrt{s}$ = 540 GeV (9). Arrow points to slope change.

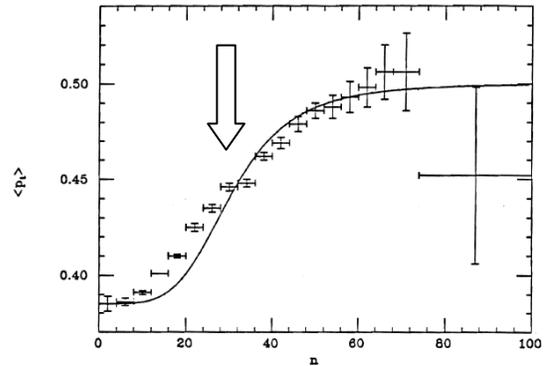

Fig. 12 Minijets model compared to UA1 data at $\sqrt{s}$ = 540 GeV(2) Arrow points to slope change.

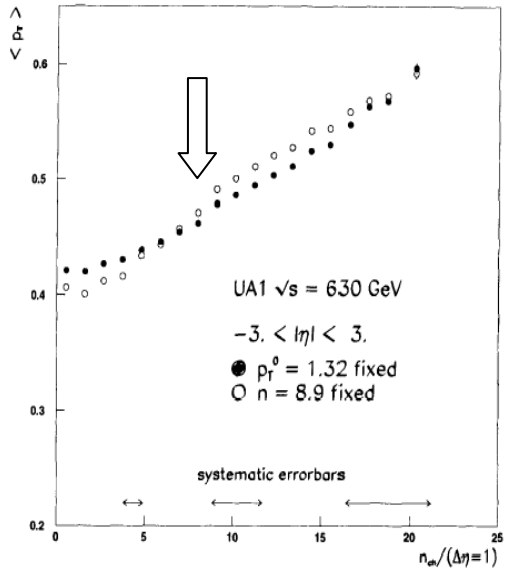

Fig. 13 <p_t> vs dn/dη , UA1 data, √s = 630 GeV (13). Arrow points to slope change.

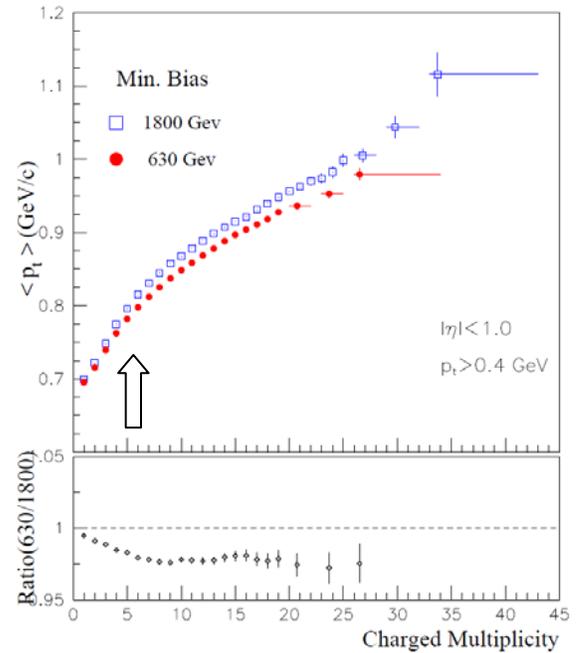

Fig. 14 <p_t> vs n , |η|<1.0, p_t >0.4 GeV/c , CDF RUN I, √s = 1800 and 630 GeV (14). Arrow points to slope change.

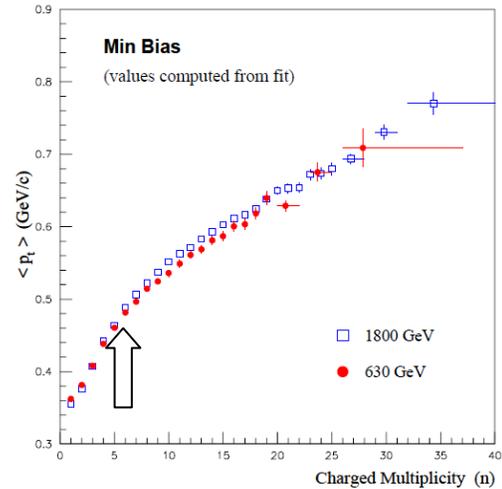

Fig. 15.As in fig.14, with p_t extrapolation to 0. Arrow points to slope change.(16)

At Tevatron the $<p_t>$ vs $\rho$ dependence has been studied by CDF and E735 collaborations.

The CDF (14; 15) results are at $\sqrt{s}$ = 630 and 1800 GeV (CDF RUN I) and at 1960 GeV (RUN II) and are shown in fig.14-20. The $<p_t>$ has been estimated from arithmetic mean within kinematical cuts $p_t > 0.4$ GeV/c and $|\eta|<1$ , for all events with a given multiplicity. The multiplicity has been measured by charged tracks number within the same kinematical space used for $<p_t>$ estimate. A plot in RUN I data has been done also with extrapolation of $p_t$ to 0 by an exponential model, see fig.15. (16),where the charged multiplicity n is still measured for $p_t > 0.4$ GeV/c and $|\eta|<1$.The data points are corrected for acceptance losses and track finding efficiency at each multiplicity. The $p_t > 0.4$ GeV/c cut gives an higher increase respect to UA1 measurements and of course higher $<p_t>$. At all 3 energies there is a slope change at multiplicities between 5 and 6. To correct multiplicity for the $p_t > 0.4$ GeV/c cut, we compare values of the average pseudorapidity densities for $|\eta| < 1$ and $| p_t | > 0.4$ GeV/c, which are 1.6 at 630 GeV and 2.1 at 1800 GeV respectively(17), to the acceptance corrected averages dn/dη , which are $3.20 \pm 5\%$ at 630 GeV and $3.95 \pm 5\%$ at 630 and 1800 GeV respectively (18). The acceptance correction factor results $3.20/1.6 = 2.0 \pm 0.2$ for 630 GeV and $3.95/2.1 = 1.9\pm0.2$, at 1800 GeV,where uncertainties on correction factors are estimated at 95% confidence level. After corrections for acceptance cuts, the multiplicities 5 and 6 correspond to dn/dη=5 and dn/dη=6 respectively, at both energies,from which we can get a nominal value of $\rho$=5 5 at the slope change.The bin size gives an uncertainty of $\pm$ 0.5,at 100% confidence level.We get a total uncertainty of $\pm0.6$ on the dn/dη at approximately 95% confidence level.

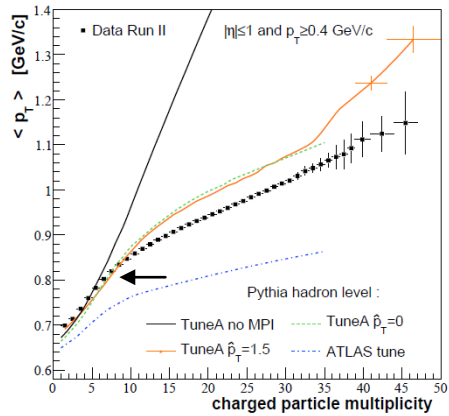

Fig. 16. <p$_t$> vs n , CDF RUN II, $\sqrt{s}$ = 1960 GeV (15). Arrow points to slope change.

In fig.17 and 18 from (16) one can see that the numerically calculated first and second derivative of the <p$_t$> vs n curve show confirm of the slope change around 6.

The 1960 GeV CDF data have been compared with models. In fig.16 one can see how the considered models don't give in general satisfactory fit.

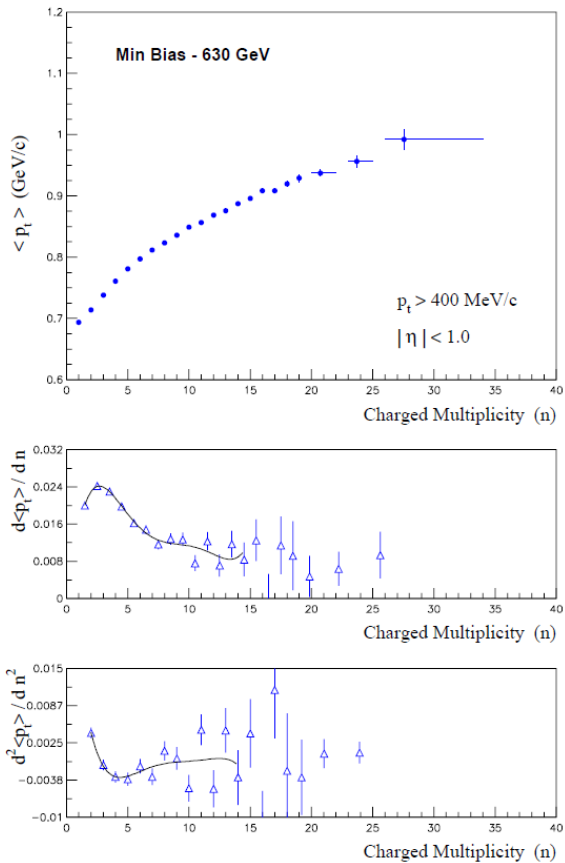

Fig. 17. <p$_t$> vs n, CDF, $\sqrt{s}$ = 630 GeV, with first and second derivatives (16)

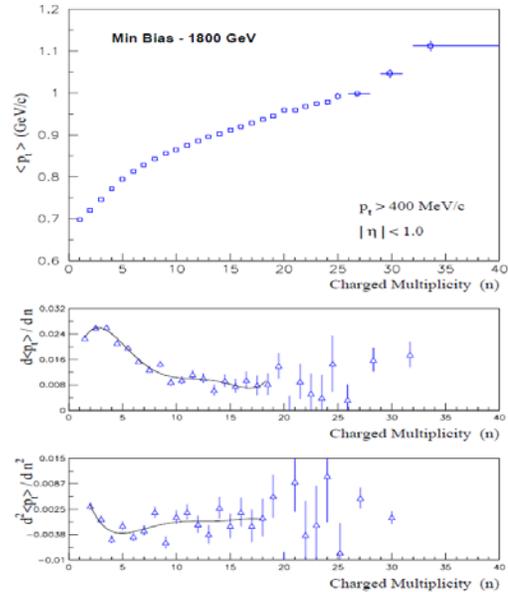

Fig. 18. <p$_t$> vs n, CDF, $\sqrt{s}$=1800 GeV, with first and second derivatives (16)

As we mentioned before, a possible interpretation of the <p$_t$> vs n increasing is in term of "minijets". According at least to some minijets model, minimum bias events contain two distinct components: a "soft" one, with a constant vs n <p$_t$> low value and an "hard" or "semi-hard" component where a hard scattering happened, with a constant <p$_t$> value, higher than the "soft" <p$_t$>. A possible third component due to many parton hard scattering can contribute. In the model, the percentage of "semi-hard" events increases with multiplicity and this fact causes the increasing of <p$_t$>. The CDF collaboration divided minimum bias events in "hard" events with a particles cluster with transverse energy E$_t$ > 1.1 GeV and "soft" events, with no such cluster. They studied p$_t$ vs n separately in "soft" and "hard" sample, both at 1800 and 630 GeV, see fig.19 and 20. We can still see a dependence of <p$_t$> on n, both for "soft" and "hard "events. The <p$_t$> vs n in "soft" events is very similar at 630 and 1800 GeV. There is still a slope change at both energies around n=6. In hard events the <p$_t$> vs n increasing is steeper compared to soft events and a weak slope change is still visible at n=6. To test results, a separation in hard and soft events has been done in the experiment with different cut E$_t$ > 3.0 GeV and results are consistent with results for E$_t$ > 1.1 GeV cut.

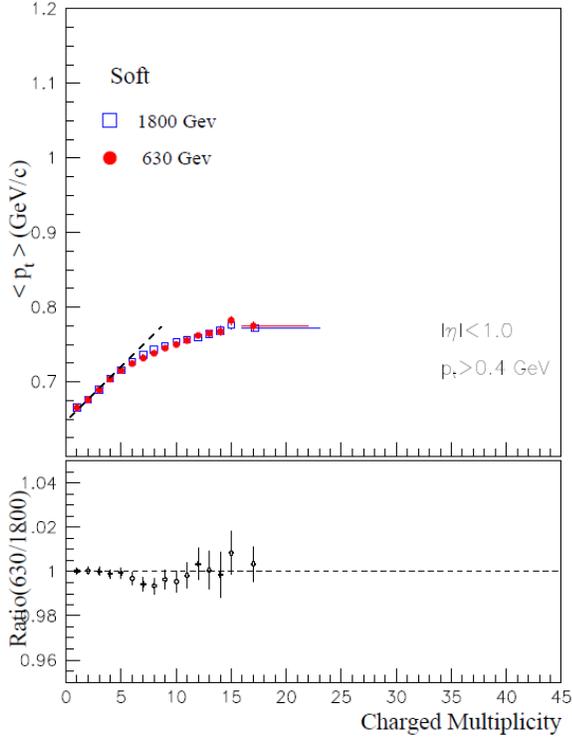

Fig. 19. <p$_t$> vs n , "soft" events (14). The shaded line must be considered as an eye guide.

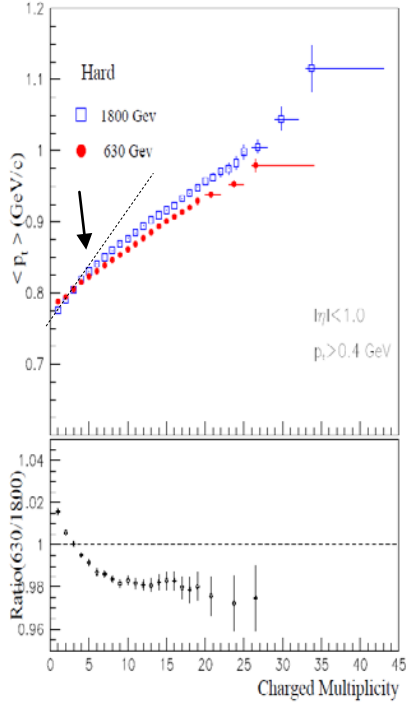

Fig. 20 <p$_t$> vs n, "hard" events (14). Arrow points to slope change. The shaded line must be considered as an eye guide for the 1800 GeV data.

A similar study has been done previously by UA1 collaboration (2). They divided events in "jetty" (with cluster E$_t$>5 GeV) and "non jetty" events (no cluster). In fig. 21, results for <p$_t$> vs n at $\sqrt{s}$ = 540 GeV are shown. The <p$_t$> vs n behavior is different between "jetty" and "non jetty" events. In the "non jetty" events a clear increasing of <p$_t$> vs n is seen and still there is the slope change at multiplicity around 28, which corresponds to $dn/d\eta$=5.6.

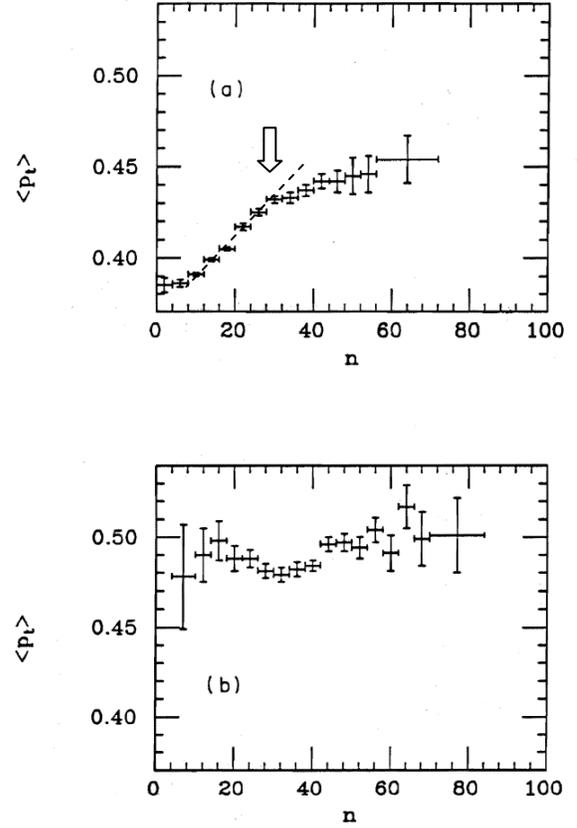

Fig. 21. <p$_t$> vs n "no jet sample" (a) and for "jet sample"(b). UA1 data, $\sqrt{s}$=540 GeV, |$\eta$|<2.5 (2). Arrow points to slope change. Shaded line must be considered as an eye guide for the "no jet sample".

The <p$_t$> rise in "soft" or "non jetty" events may be explained by the fact that not all "jetty" or "hard" events have been removed from the "soft" sample by the E$_t$ cuts. However the E$_t$ cut of 1.1 GeV at CDF is not a high energy cut. In (19), a study has been done on the structure of minimum bias events selected for the presence of a particle (chjet#1) with p$_t$ > 2 GeV/c (fig.22, 23). The charged particle distribution as function of the difference in azimuthal angle to the chjet#1 particle direction has been studied. We see in fig.23 that particle distribution does not present the typical back to back structure of jet events. That may be due to the fact that the |$\eta$|<1 acceptance is small and back jet particles are not detected. If we look at minijets models (20), we see in fig.24 that , in case of a limited rapidity |y|<1 acceptance, they predict higher particles density in the "away" zone in comparison to the "transverse" one, even for chjet#1 particle p$_t$ >1.0 GeV/c.

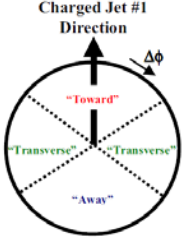

Fig. 22. Toward,Transverse and Away azimuthal region definition respect to Charged Jet #1 Direction (19).

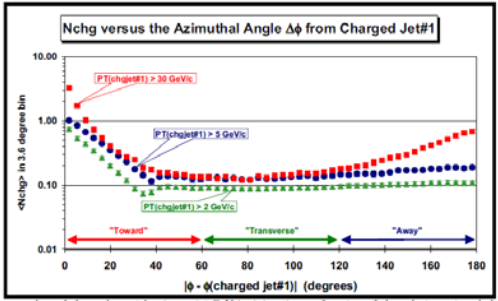

Fig. 23 Charged particle number versus the Azimuthal Angle $\Delta\phi$ from Charged Jet#1 for different Charged Jet#1 $p_t$ (19).

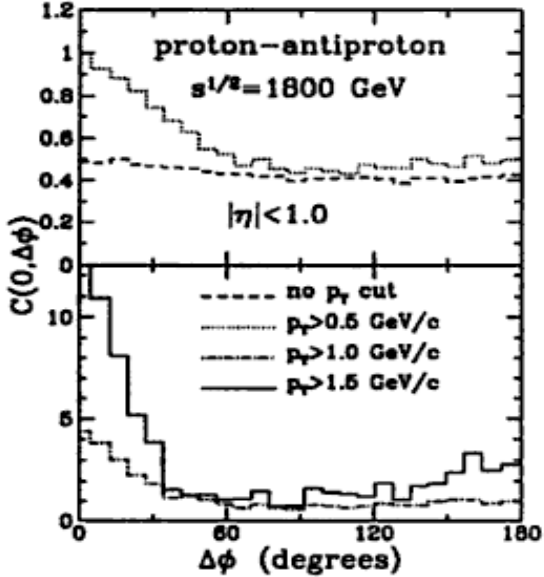

Fig. 24 HIJING model prediction for Charged particle number versus Azimuthal Angle $\Delta\phi$ from jet direction, for |y|<1 acceptance, for various $p_t$ cut in jet leading particle transverse momentum (20).

These results question the fact that the $<p_t>$ rise is completely due to the increasing of percentage of "jetty" events with multiplicity and that the "hard" or "jetty" events in minimum bias events with low to medium transverse energy are all really hard scattering events.

In (21), events from STAR collaboration experiment at RHIC in $pp$ at $\sqrt{s} = 200$ GeV were subdivided in "hard" and "soft" events, as in CDF, with $E_t > 1.2$ GeV. In fig.25 from (21), results are plotted together with CDF results. The $<p_t>$ has been calculated for $p_t > 0.4$ GeV/c. For the pseudorapidity acceptance difference between detectors, HIJING (12) Monte Carlo has been used to extend the acceptance of STAR from $|\eta|<0.5$ to $|\eta|<1.0$. The pseudorapidity bin size in STAR is about a factor 1.5 greater than the bin size in CDF. The STAR $<p_t>$ increases with multiplicity in the minimum bias with a week slope change at $N_{ch} \approx 8$.It increases in soft events as well, in a similar way than in CDF.

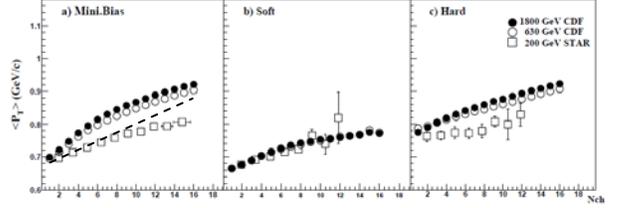

Fig. 25 The mean transverse momentum $<p_t>$ dependence on the multiplicity $N_{ch}$ in minimum bias,soft and hard events. The STAR acceptance is extended from $|\eta|<0.5$ to $|\eta|<1.0$ by HIJING Monte Carlo to compare the results with CDF. From (21).Shaded line is only for reader eye guiding.

More recently, the STAR collaboration analyzed the data with a model based on hard and soft components (22). Fit has been done to find the best parameters of $p_t$ distributions at different multiplicities, the best values of $<p_t>$ and of the percentages of soft and hard component at each multiplicities. An analysis of residual at each observed multiplicity is presented. STAR contrasts own analysis with the analysis based on power law fit at each multiplicity. The residual between experimental and fitted values for the power law analysis is presented in fig.26 as function of transverse rapidity $y_t = \ln\left(\frac{m_t + p_t}{m_0}\right)$, with transverse mass $m_t \equiv \sqrt{p_t^2 + m_0^2}$ and pion mass $m_\pi$ assumed for $m_0$, at various observed multiplicity $\hat{n}_{ch}$. The residual are quite big at all multiplicities and at all $y_t$ by using power law fit.

Double component STAR fit procedure (fig 27) residuals look good at high multiplicity in the full $y_t$ range, but not in the low $y_t$ at multiplicity $\leq 5$. In fig.28 the $<p_t>$ vs n dependence obtained by the two component fit to STAR data is linear in the observed, not corrected, multiplicity $\hat{n}_{ch}$.The hatched region in fig.28 represents the common uncertainty in all means due to uncertainty in the particle yield in $p_t<0.2$ GeV/c. These uncertainties are quite large and particularly important for $\hat{n}_{ch} \leq 4$, which, corrected for tracking inefficiencies and $p_t <0.2$ GeV/c cut, corresponds roughly to $dn/d\eta \leq 8$. Because residuals are not good at low $y_t$ at low multiplicities, it may be that the two component model does not give a good representation of the soft physics. The STAR analysis aim is to provide isolation of the hard component of the $p_t$ spectrum as a distribution of simple form on $y_t$. In (23) authors assert "that both soft and hard parts presented in (22) can have very large transverse momentum.This in contradiction with our normal picture about the soft part which is always assumed having small transverse momentum".

In the power law fit results to STAR data one can see a slope change at the not corrected multiplicity $\hat{n}_{ch} \approx 4$ (fig.28).

Due to uncertainty in STAR data analysis results, we can not reach any conclusion on the existence of slope change in $pp$ STAR data.

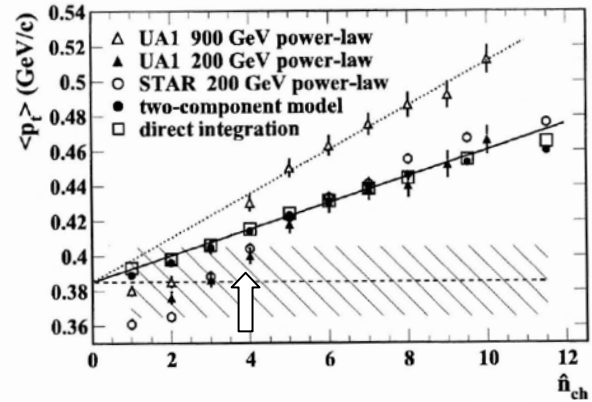

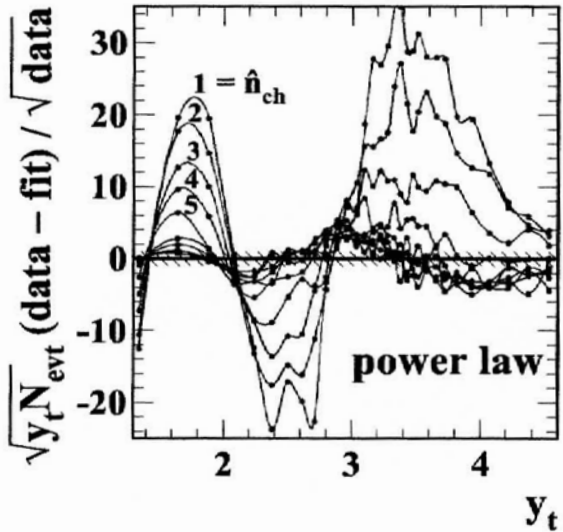

Fig. 26.Residuals of power law model fit to data as function of observed $\hat{n}_{ch}$ and tranverse rapidity $y_t$ (22)

Fig. 28 Results of STAR coll. fits to $p_t$ data . $<p_t>$ ($\hat{n}_{ch}$) derived from the two–component Gaussian amplitudes (solid dots), from the running integrals (open squares) and from power law fits to STAR and UA1 data (open circles, triangles) (22). The hatched region represents the common uncertainty in all means due to uncertainty in the particle yield in $p_t < 0.2$ GeV/c. The two component and the direct integration model fit data in a linear way. The arrow points to the slope change in the power-law fit results.

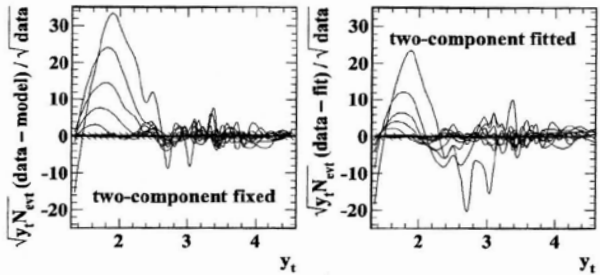

Fig. 27 Residuals of STAR fit as function of observed multiplicity $\hat{n}_{ch}$ and transverse rapidity $y_t$(22).

One may object that the slope change existence depends on the applied analysis method. Actually in the ISR and CDF experiments the $<p_t>$'s have been calculated by arithmetic mean, which is a model free procedure. In other experiments ( E735,ALICE, CMS) the arithmetic mean has been used as well and in their results we will see the slope change.

The $<p_t>$ vs $\rho$ dependence has been studied also by E735 (24) collaboration at TEVATRON at 1800 GeV. In their experiment the $<p_t>$ value has been calculated by average of $p_t$ values for $0.150 < p_t < 3.0$ GeV/c and $-0.36 < \eta < 1.00$. In the case of negative tracks, results are presented also with extrapolation by exponential law to $p_t < 0.150$ GeV/c. The $\rho$ density in pseudorapidity has been obtained dividing by 6.5 the multiplicity for $|\eta| < 3.25$. Data were corrected for acceptance .In fig.29 we see a slope change at $dN_c / d\eta$, roughly at $5.5 \pm 0.7$, with uncertainty given by binning size.

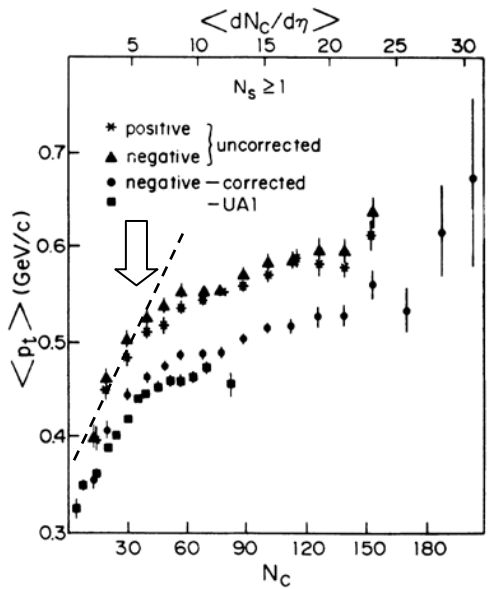

Fig. 29 $<p_t>$ vs multiplicity $N_c$ and $<dN_c/d\eta>$ from experiment E735 at 1800 GeV for all positive and negative tracks averaged over the interval 0.150 GeV/c $< p_t <$ 3.0 GeV/c. The corrected $<p_t>$ of negative particles includes unobserved particles with $p_t < 0.150$ GeV/c. Arrow points to slope change. Dashed line must be considered as an eye guide (24).

In (25), the results of ALICE experiment at LHC in $pp$ collisions at $\sqrt{s} = 900$ GeV are presented. In fig.30 we see the $<p_t>$ vs n, with and without acceptance corrections. Results for $<p_t>$ calculated by arithmetic mean for $p_{t\,min} > 0.500$ GeV/c, $p_{t\,min} > 0.150$ GeV/c and by extrapolation to $p_t = 0$ are presented. The multiplicity is measured for $|\eta| < 0.8$. One can see a slope change at multiplicity between 9 and 10 in the not corrected data and between 10 and 11 in the corrected ones, which corresponds to dn/d$\eta$ = 6.5 for $|\eta| < 0.8$. The uncertainty on the dn/d$\eta$ is mainly due to bin size($\pm$0.3) and to the determination of the slope change point($\pm$0.3). We estimate total uncertainty to be 0.5 at 95% confidence level. In fig. 31, one can see that the Pythia Perugia model fits well the data with $p_t > 0.5$ GeV/c, but not with $p_t > 0.1$ GeV/c. Other models fail to fit with both $p_{t\,min}$.

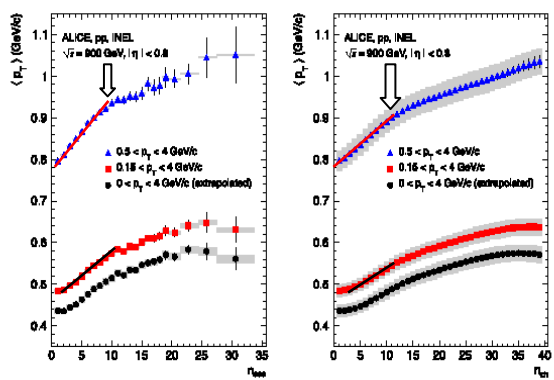

Fig. 30. $<p_t>$ vs raw $n_{acc}$ and corrected $n_{ch}$ multiplicity (25). Arrows point to $<p_t>$ vs n slope change multiplicity. Lines are only for reader eye guiding.

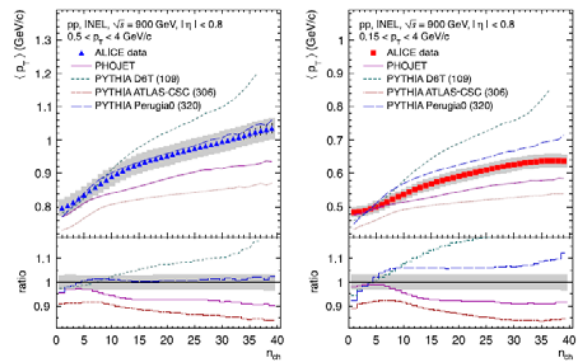

Fig. 31. The average transverse momentum of charged particles for 0.5$< p_t <$4 GeV/c (left panel) and 0.15$< p_t <$ 4 GeV/c (right panel) at $\sqrt{s} = 900$ GeV as function of corrected multiplicity $n_{ch}$ in comparison to models (25).

At the same energy of 0.9 TeV in $pp$ collisions, ATLAS (26) collaboration presented the results shown in fig. 32, where we see a slope change at $n_{ch}$ about 10. The measurement has been done in -2.5 < $\eta$ < 2.5, with $p_{t\,min} > 0.5$ GeV/c. We need to correct the multiplicity for the $p_t$ acceptance cut to compare with other experiments. From (26), the pseudorapidity density dn/d$\eta$ is $1.333 \pm 0.003(stat.) \pm 0.040(syst.)$ in the $\eta$ central region with $p_t > 0.5$ GeV/c. ALICE experiment(27) at the same energy found an acceptance corrected value of dn/d$\eta$ = $3.81 \pm 0.01 \pm 0.01(syst.)$ : this gives a ratio of 3.81/1.333 = 2.86, which we use to obtain the dn/d$\eta$ for $p_t > 0$. The slope change is seen at about n=10, which corresponds to dn/d$\eta$= (10x2.86)/5 = 5.72. The uncertainty on dn/d$\eta$ is due to bin size ( $\pm$ 0.2) ,acceptance correction ($\pm$ 3%) ,slope change point determination ($\pm$ 0.1). We can estimate a total uncertainty of 0.4 at 95% confidence level. In fig .32 we see that the Pythia Perugia model fits well the $p_t$ vs n dependence. It fits in a good way data for $p_{t\,min} > 0.5$ GeV/c as in ATLAS as in ALICE. Conversely the model does not so well for data with lower $p_{t\,min}$ cut. This may be due to a not so good low $p_t$ physics modeling. Pythia Perugia model (28) has been tuned on Tevatron minimum bias data at 630, 1800 and 1960 GeV as well SPS Collider data at 200, 540 and 900 GeV. In particular it has been tuned to the $<p_t>$ vs n at CDF RUN II with $p_{t\,min} > 0.4$ GeV/c. In fig.33 ATLAS data points are compared to EPOS model (29) simulations with hydrodynamic evolution. The slope change is nicely predicted. In EPOS model miniplasma systems may be produced in $pp$ collisions. It is important to see if the model can predict the slope changes at the different energies.

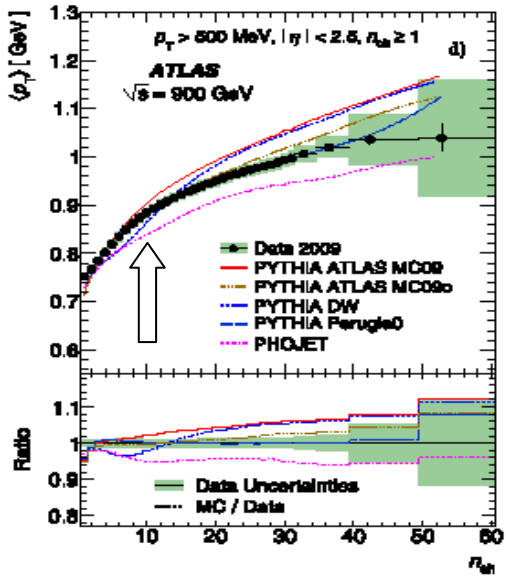

Fig. 32. The average transverse momentum as a function of the number of charged particles in the event, ATLAS (26). Arrow points to slope change.

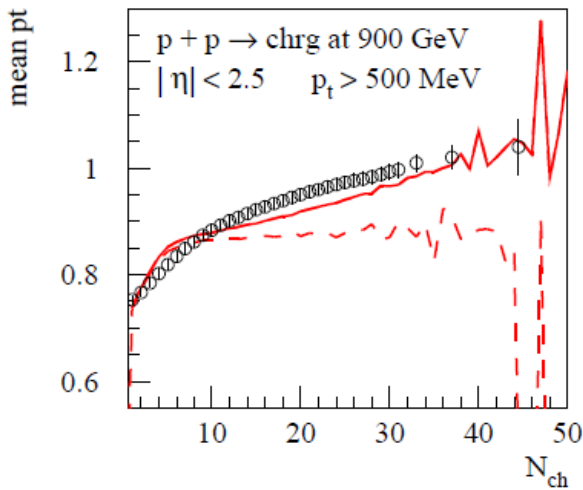

Fig. 33 ATLAS data points are compared to EPOS simulations with (full line) or without (dashed line)"hydroevolution". From (29).

from data at multiplicity near to the slope change at all the three energies. At that same multiplicity PHOBOS begins to underestimate $< p_t >$ at 0.9 and 2.36 TeV.PYTHIA 8 nicely reproduces the data at 0.9 and 2.36 Tev but at 7 TeV it overestimates $< p_t >$ beginning at n$\approx$ 27. A comment is in order: on the CMS $< p_t >$ vs n data at the three energies, 6 out of 9 models predictions go away from data in the slope change region.

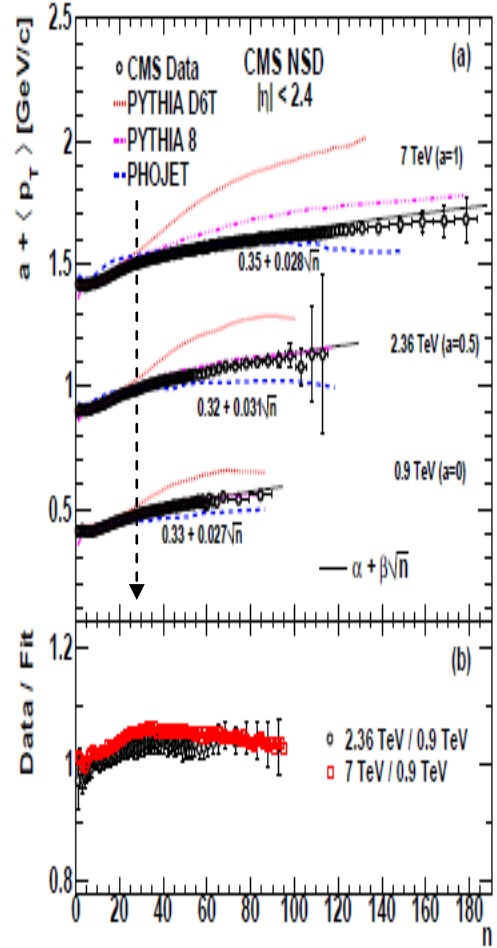

Fig. 34. A comparison of $< p_t >$ versus n for $|\eta| < 2.4$ with two different PYTHIA models and the PHOHET $\sqrt{s}$=0.9, 2.36 and 7 TeV, CMS collaboration (30). Dashed line show slope changes at various energies.

In fig .34 we present results of CMS experiment (30) for $< p_t >$ vs n at 0.9, 2.36 and 7 TeV. The $< p_t >$ and multiplicity n are corrected for acceptance. The η interval is -2.4 < η < 2.4. A slope change at the three energies is present for multiplicity between 25 and 30, which corresponds to ρ between 5.2 and 6.3, let's say $\rho = 5.7 \pm 0.6$,with uncertainty due to slope change determination.An additional uncertainty source is the bin size ($\pm$ 0.2). We estimate a total uncertainty of $\pm$ 0.8 at approximately 95% confidence level. A linear fit in $\sqrt{n}$ is shown to the data: one can see that the ratio of data at 7 TeV and at 2.36 TeV to the fit at 0.9 TeV is near to 1, but it is not flat.The fit is applied to the n>15 interval. Some models predictions are presented. PYTHIA D6T curve goes away

We saw that to compare the $< p_t >$ vs ρ dependencies at various experiments, one needs to take care of the following points:

- Different $p_t$ cuts to calculate $< p_t >$;
- Different ways to calculate $< p_t >$: arithmetic mean, data fitting, with or without extrapolation to 0;
- Different intervals of rapidity or pseudorapidity are used to calculate ρ and $< p_t >$;
- Sometimes multiplicities are corrected for acceptance cuts, sometimes they are not.This is a very important

point. The correction for acceptance is needed to compare data from different experiments and data with model predictions.To get the corrected $<p_t>$ vs $\rho$, one has to do the multiplicity correction, which is an inverse problem whose solution can smear out the true $<p_t>$ vs n dependence.

To conclude this section: in many experiments at energies from 22 GeV to 7 TeV there is a slope change in the $<p_t>$ vs $\rho$ dependence for charged particles. We showed the slope change in a total of 21 out of 22 curves: 1 curve at 22 GeV, 6 from 31 to 63 GeV, 8 from 200 to 900 GeV, 6 from 1.8 to 7 TeV. We counted only the number of independent curves: where results have been presented for positive, negative and all charged particles, we considered only positive and negative. Curves which differ from one another in $p_t$ or pseudorapidity cuts are counted once.

The only curve where slope change existence is in doubt comes from STAR experiment, as we discussed before.

Slope changes appear at $\rho$ values from $5.0 \pm 0.5$ up to $7.6 \pm 1.0$, with average 5.6, weighted average 5.5, median 5.5, and standard deviation 0.6. Nineteen out of the 21 values are in the dn/d$\eta$ interval from 5.0 to 5.7. We can say that at approximately 95% of confidence level there is slope change in the $<p_t>$ vs $\rho$ dependence at $\rho^* = 5.5 \pm 1.2$. An interesting question is if this slope change is unique in the correlation of $<p_t>$ with multiplicity n.In other words: are there other slope changes or anomalies? In $<p_t>$ vs $\rho$ correlation with no $p_{t\,min}$ cut or with low $p_{t\,min}$ cut one can see that $<p_t>$ is constant for few points up to $\rho \approx 2.0$ where it begins to increase. This slope change at $\rho \approx 2.0$ is well predicted by minijets model in 63 GeV ISR data (fig.6), in UA1 data at 200, 540, 900 GeV (fig. 8 to 12) and by PYTHIA models in CMS data at 0.9, 2.36 and 7.0 TeV (fig. 34). We are not aware of any other anomaly in the $<p_t>$ vs $\rho$ curve. Therefore we can say that the slope change at $\rho^*$ is unique in the sense that it is not explained by current models. It is interesting to note that in all curves,at all energies, the rise is steeper for $\rho < \rho^*$ than for $\rho > \rho^*$. The first part of the curve up to $\rho^*$ is well described in (1) for the 540 GeV UA1 data by decreasing impact parameter with increasing event multiplicity. LHC data seem not to confirm this model, which predicts that for a given $\rho$, the $<p_t>$ will decrease with s.The impact parameter model does not predict the slope change at $\rho^*$.

Minijets models(9) describe quite well the $<p_t>$ raising with $\rho$, but they don't reproduce the sudden slope change. They predict at $\sqrt{s} > 1800$ GeV the so called "ledge" effect at $\rho$ values from $\approx 10$ up to 30, but Tevatron and LHC data do not show this effect.Moreover ,at the highest energies minijets models produce too many high $p_t$ particles(31).

Even PITHYA 8 and PITHYA D6T models at the highest energies seem to produce too high $p_t$ particles and begin to be at variance with data at $\rho^*$. On the contrary, PHOJET underestimates the rise .

PITHYA PERUGIA describes well data with $p_{t\,min}$ cut >0.4 GeV/c but not data with lower $p_{t\,min}$ cut.

The EPOS model by miniplasma production and hydrodynamic evolution predicts the slope change in ATLAS data and also the behavior at low and high $\rho$ , but we don't know yet if it reproduces the slope change in different data.

The thermodinamical model(3) suggested, as a possible signal for the deconfinement transition of hadronic matter, the flattening in the early UA1 $<p_t>$ curve for $\rho$ values > 8. Data at higher energy (Tevatron, LHC) do not confirm this plateau at high $\rho$ values. Even in high statistics UA1 data at 630 GeV there is no evidence of saturation of the increase over the measured range up to $\rho = 20$.

In conclusion ,it seems that the slope change at $\rho^*$ for so many energies is an "anomaly" at variance with all models.

## III. CORRELATED SIGNALS

It is very important to check if the events with $\rho$ values near to the slope change value $\rho^*$ have characteristic features. It may be very interesting even to look at possible differences between events with $\rho$ lower or higher than $\rho^*$. This task can be done with data from a single experiment or from many experiments. We considered some phenomena that may be important to this aim. For some of them there is a satisfactory number of experimental results, while for the others only very few data exist and we will mention them just to suggest possible experimental investigations.

Here we list some of these interesting signals:

- $<p_t>$ vs $\rho$ for identified particles;
- strange particles production as function of $p_t$ and $\rho$;
- Multiplicity distributions and KNO violation;
- Source dimension dependence on $\rho$ and on the transverse momentum $k_t$ of particle pairs;
- Fluctuations event by event of $<p_t>_{ev}$ as function of $\rho$;
- Short and long range rapidity particles correlations as function of $\rho$ and of $<p_t>_{ev}$;
- Spike events ;
- Low $p_t$ positron production as function of $\rho$;
- Mass shift of the $\rho^0(770)$ resonance and its multiplicity dependence.

In case of sufficient experimental resolution in multiplicity or $\rho$ density we look to details in the slope change region, otherwise we check if data are consistent with the hypothesis of difference between events with $\rho$ density higher or lower than $\rho^*$.

### A. $<p_t>$ DEPENDENCE ON $\rho$ FOR IDENTIFIED PARTICLES AND STRANGE PARTICLES PRODUCTION.

$<p_t>$ vs n for identified particles has been studied among others by E735 collaboration (32). In E735 the $p_t$ vs $\rho$ dependencies have been studied for $\pi$, k, $\bar{p}$, $\Lambda_0$, and $k_0$ and other strange particles. The $<p_t>$ value has been obtained by power law fit for $\pi$ and exponential fit for k , $\bar{p}$ ,from the measured values in 0.150 GeV/c< $p_t$ < 1.5 GeV/c and in -0.36<$\eta$<1.0. The pseudorapidity density $dN_c/d\eta$ has been calculated dividing by 6.5 the multiplicity in -3.25< $\eta$ < 3.25. The result for charged pions and kaons and for antiprotons are presented in fig 35, for 0< $p_t$ < 1.5 GeV/c,with pseudorapidity density bin size of 1.4 unit. One can see the slope change at $\rho = 5.7 \pm 0.7$ in $<p_t>$ vs $\rho$ in all curves. We used the bin size as uncertainty measure.

We note that these results are not independent from the ones previously shown for negative and positive charged particles for the same experiment: the $<p_t>$ vs $\rho$ for all charged particles is a weighted combination of the $<p_t>$ vs n for the identified charged particles. In fig.36 the $\bar{p}$ results are shown together with the $\Lambda_0$ and $\overline{\Lambda_0}$, for which there are only few points. The $\Lambda_0$ and $\overline{\Lambda_0}$ spectra have been measured in the 0.5-3.0 GeV/c range and extrapolated by exponential law to $0 < p_t <$ infinity.

The $\bar{p}$ $<p_t>$ increasing rate is nearly equal to that of $(\Lambda_0 + \overline{\Lambda_0})$. There is a kind of step in the antiproton graph at dn/dy about $7.0 \pm 1.0$, with uncertainty due to bin size. The most recent $\bar{p}$ results with highest statistics from E735 have been shown in fig.35.

It is worthwhile to note in fig.35 that the $<p_t>$ increase is much faster for heavier particles and that the difference among the $<p_t>$'s for the different particles gets bigger after the slope change region, especially for $\pi$ in comparison to $\bar{p}$ and to k. These results are qualitatively in agreement with models which consider phase transition from hadronic matter to Quark Gluon Plasma (fig. 37) (33) and we will return on them and on similar results from other experiments.

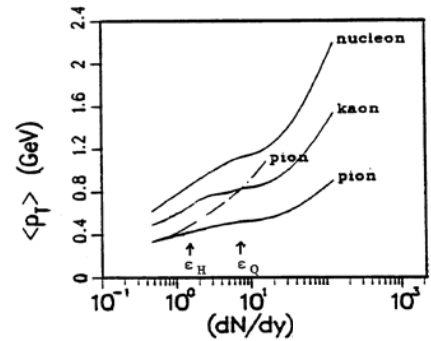

Fig. 37 $<p_t>$ vs dn/dy for pion without phase transition ( shaded line) and for pion, kaon and nucleon with phase transition (33).

The results on $<p_t>$ vs $\rho$ for neutral particles give us information which is new respect to the behavior of all charged particles.

Besides E735, the CDF, UA1 and STAR collaborations among others studied $k_0$ and $\Lambda_0 <p_t>$ vs n dependence.

UA1(34) results are shown in fig.38. The $<p_t>$ rise is much higher for $\Lambda_0$ than for $k_0$ and for charged hadrons. Data points are too sparse to look for possible structure in $<p_t>$ vs $\rho$ dependence.

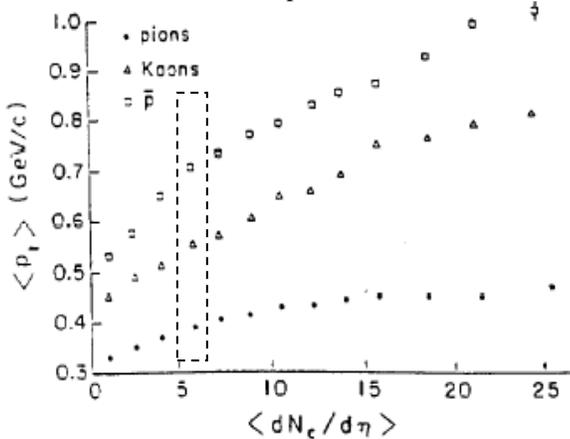

Fig. 35.Average transverse momentum$<p_t>$ as a function of charged particle density $<dNc/d\eta>$ from E735 coll. (32). Box shows slope change region.

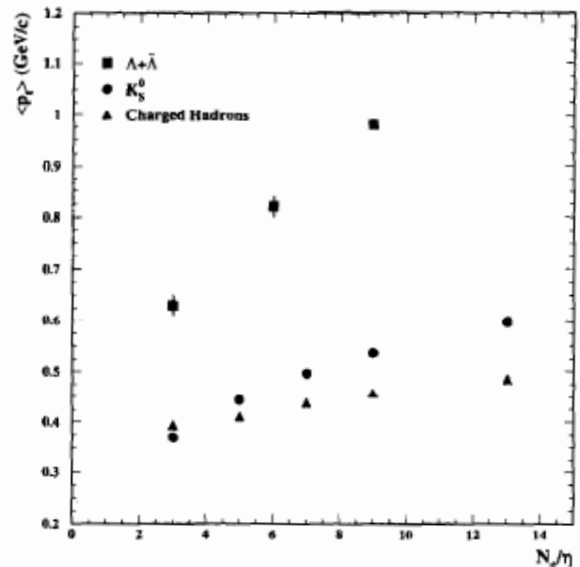

Fig. 38 Average $p_t$ as a function of charged track multiplicity per unit pseudorapidity for charged hadrons, $k_0$, and $(\Lambda_0 + \overline{\Lambda_0})$ by UA1 collaboration (34).

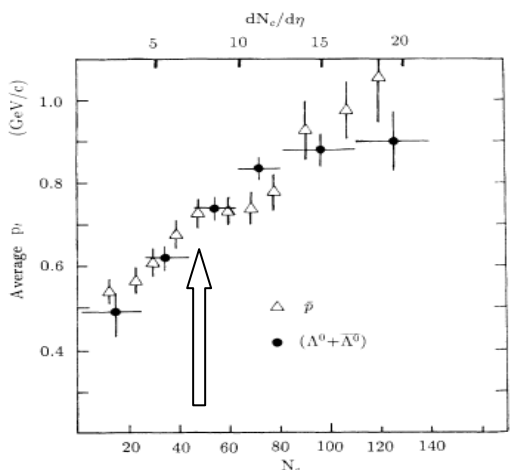

Fig. 36.Average transverse momentum $<p_t>$ vs charged multiplicity $N_{ch}$, from E735 coll. (32)Arrow points to slope change.

CDF studied the $<p_t>$ vs n for $\Lambda_0$ and $k_0$ at 630 and 1800 GeV, in minimum bias events (35). As for the charged $<p_t>$ study the minimum bias events have been split up into "soft" and "hard" event subsamples. In Fig.39 to 44 the $<p_t>$ vs the charged particle multiplicity n are shown for minimum bias, soft and hard events at the two energies. The $\Lambda_0$ and $k_0$ $<p_t>$'s have been calculated with $p_{t\,min}$ cut $> 0.4$ GeV/c. The charged particles multiplicity is measured in $p_t > 0.4$ GeV/c and $|\eta| < 0.5$ phase space, as in the charged $<p_t>$ case. Data are corrected for experimental inefficiencies. One can see in the minimum bias events an increase with multiplicity for both $\Lambda_0$ and $k_0$ $<p_t>$. At n around 6 there is a clear slope change in kaons curve and, less evident, for lambdas, both at 1800 and 630 GeV. An increase of the mean $p_t$ is observed also in the "soft" subsample alone, a feature that is not explained by the current models. We note that in minimum bias from n=6 onward, $<p_t>$ is increasing much faster for lambdas in comparison to kaons and to charged particles. Fig.39 and 40 show a fan shaped behavior which widens for n>6, where the charged $<p_t>$ goes away from kaons $<p_t>$ and, more strongly, it goes away from lambdas $<p_t>$. In "hard" events, the $<p_t>$ for kaons and lambdas are quite similar up to n around 6, but for n > 6 the $<p_t>$ increasing rate is much higher for lambdas than for kaons. In "soft" events, instead, the increases for lambdas and kaons are quite similar and both show a slow down for n about 6, both at 630 and 1800 GeV. This means that the increase of $<p_t>$ for $n > 6$ both for kaons and lambdas in minimum bias events is due only to the increase in "hard" events.

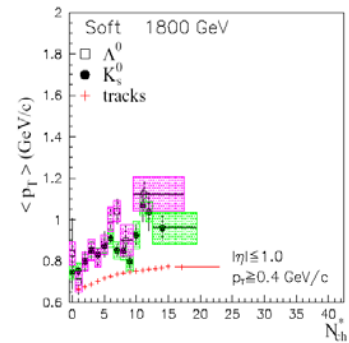

Fig. 41 $<p_t>$ vs $N_{ch}$ for $k_0$, $\Lambda_0$ and charged particles, in "soft" events $\sqrt{s} = 1800$ GeV, CDF (35).

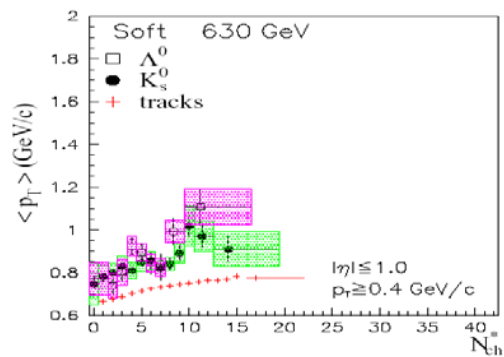

Fig. 42 $<p_t>$ vs $N_{ch}$ for $k_0$, $\Lambda_0$ and charged particles in "soft" events, $\sqrt{s} = 630$ GeV, CDF (35).

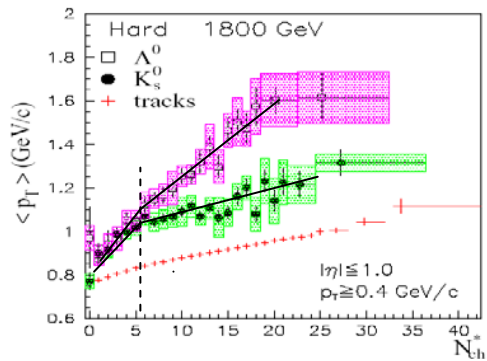

Fig. 43 $<p_t>$ vs $N_{ch}$ for $k_0$, $\Lambda_0$ and charged particles in hard events, $\sqrt{s} = 1800$ GeV, CDF (35).Full lines are only for reader eye guiding..Dashed line is drawn in correspondence of slope change.

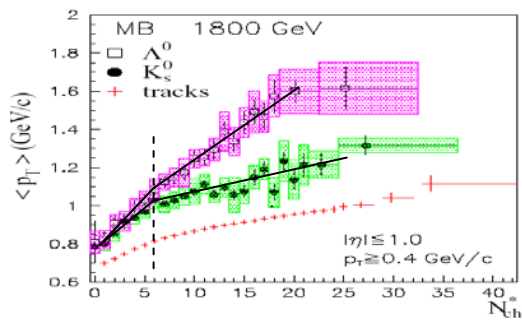

Fig. 39 $<p_t>$ vs charged multiplicity $N_{ch}$ for $k_0$, $\Lambda_0$ and charged particles, in minimum bias events, $\sqrt{s} = 1800$ GeV, from (35). Full lines are only for reader eye guiding. Dashed line is drawn in correspondence of slope change.

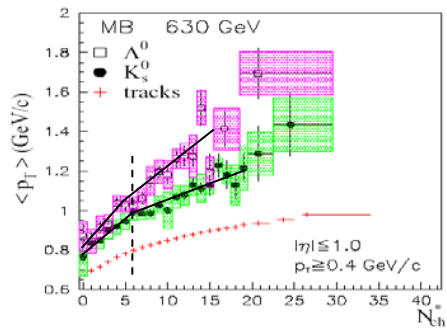

Fig. 40 $<p_t>$ vs $N_{ch}$ for $k_0$, $\Lambda_0$ and charged particles, minimum bias, $\sqrt{s} = 630$ GeV. CDF(35).Dashed line is drawn in correspondence of slope change.Full lines are only for reader eye guiding.

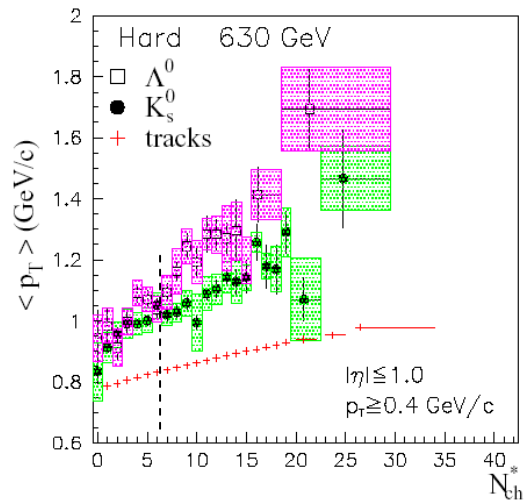

Fig. 44 $<p_t>$ vs $N_{ch}$ for $k_0$, $\Lambda_0$ and charged particles in hard events, $\sqrt{s} = 630$ GeV, CDF(35). Dashed line is drawn in correspondence of slope change.

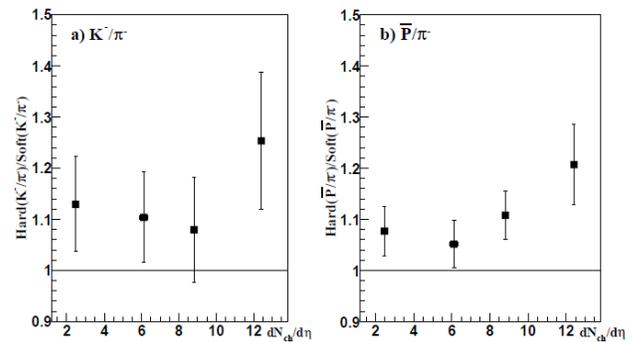

Fig. 46 The distributions of hard($K^-/\pi^-$)/soft($K^-/\pi^-$), hard($\bar{p}/\pi^-$)/soft($\bar{p}/\pi^-$) in the midrapidity ($|y|<0.2$). The error bars are systematic errors.(21)

From the inclusive $p_t$ distributions for $\Lambda_0$ and $k_0$ from CDF, UA1 (fig.47) and STAR (fig.48) collaborations at 1800, 630 and 200 GeV respectively, one can get the $\Lambda_0/k_0$ ratio as function of $p_t$ (36; 37), together with models predictions. One can see that for $p_t > 0.5$ up to about 4.0 GeV/c, the $\Lambda_0/k_0$ ratio increases and that for $p_t > 1$ GeV/c no one of the shown models can explain this high ratio.

Some comments are in order. From CDF (35) we get that the fractions of total $\Lambda_0(k_0)$ that falls into the hard subsample are quite high, ranging from about 85% (70%) at 630 GeV to about 90%( 82%) at 1800 GeV. The percentage of "hard" events with at least one $k_0$ is about 13% and with at least one $\Lambda_0$ is about 7%, both at 1800 and 630 GeV. These percentages go down to 4% and 1% respectively in "soft" events, at both energies. The percentage of "hard" events with $\Lambda_0$ is seven fold the correspondent percentage of "soft" events. The ratio ($\Lambda_0$ events)/( $k_0$ events) is much higher in "hard" sample than in "soft" sample and than in minimum bias. The mean number of $\Lambda_0$ per event divided by the charged multiplicity is more than a factor two higher in the "hard" sample than in the soft one. In (21), authors found that in STAR experiment data in pp at $\sqrt{s}$=200 GeV the $k^-/\pi^-$ and $\bar{p}/\pi^-$ratio in "hard" events were higher than in "soft" events and that at $<p> \cong 12$ their values were nearly equal to the values measured in the Au-Au central collisions at RHIC (fig.45, 46).

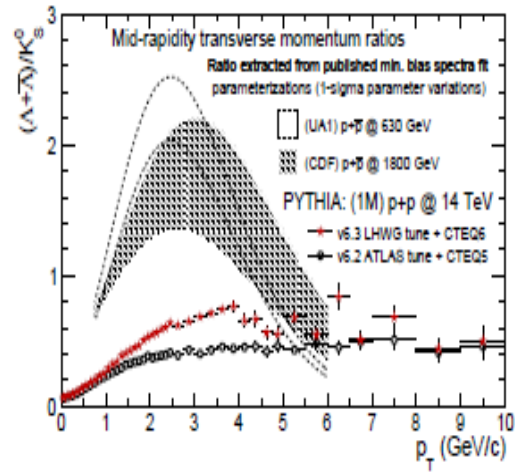

Fig. 47 $(\Lambda_0 + \overline{\Lambda_0})$ / $k_0$ ratio as a function of $p_t$ from UA1 and CDF collaborations. Data are compared to model prediction (36).

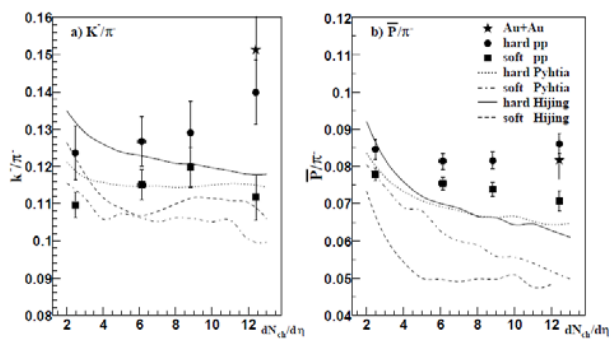

Fig. 45 $K^-/\pi^-$ and $\bar{p}/\pi^-$ ratio versus $\frac{dN_{ch}}{d\eta}$, in $pp$ $\sqrt{s} = 200$ GeV and in Au Au collisions. Lines are models predictions. See ref. (21).

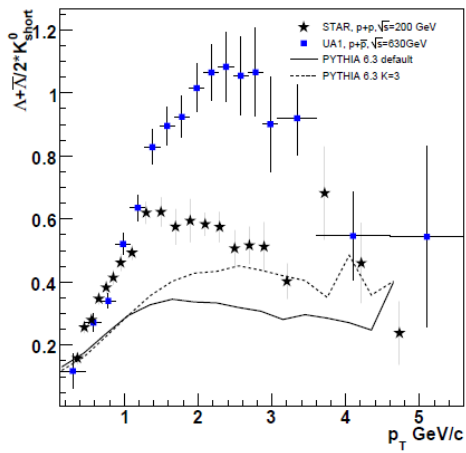

Fig. 48  $(\Lambda_0 + \overline{\Lambda_0})/(2 * k_0)$ ratio as a function of $p_t$ from STAR collaboration (37).

In fig. 49 the $\Lambda^0/K^0$ ratios from STAR collaboration (37) in $pp$ at 200 GeV are shown for three $<dN_{ch}/d\eta>$ mean values: one can see that roughly for $p_t > 1.3$ GeV/c the ratio increases with $<dN_{ch}/d\eta>$. We note that the ratios at $<dN_{ch}/d\eta> = 4.68$ and at $<dN_{ch}/d\eta> = 9.01$ are similar and that they are different to the ratio at low $<dN_{ch}/d\eta> = 1.98$. This dependence of the $\Lambda^0/K^0$ ratio on $<dN_{ch}/d\eta>$ in $pp$ collisions is similar to the dependence on events centrality in Au-Au at RHIC at 62.4 GeV (38) (fig.50).

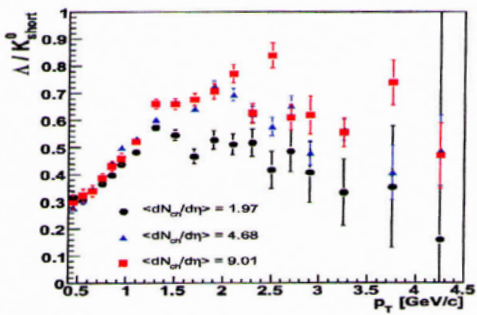

Fig. 49.  $\Lambda^0/K^0$ ratio as a function of transverse momentum for three average multiplicity bin in $pp$ collisions at $\sqrt{s} = 200$ GeV. STAR coll. (37).

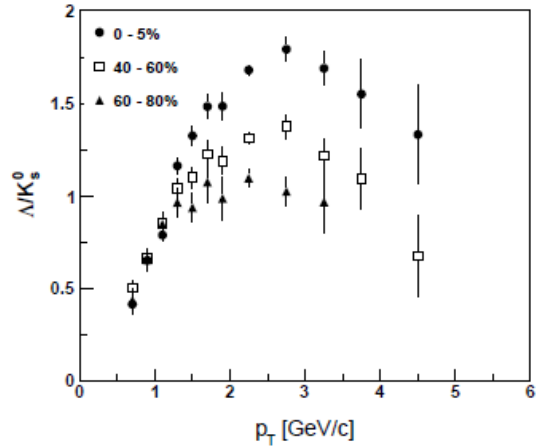

Fig. 50.The $\Lambda^0/K^0$ ratio as a function of transverse momentum for different centrality classes in Au-Au collisions at 62.4 GeV. The most central class is the 0-5% (38).

Recently minijets model HIJNG/BB-v2.0 (31) well described the large strange baryon-to-meson ratios at Tevatron energies. At 7 TeV the model gives a good description of CMS $(\Lambda_0 + \overline{\Lambda_0})$ production, but it underestimates by approximately a factor of two the yield of $k_0$'s and, as a consequence, it overestimates the ratio. Besides in the model there is an overestimation of high $p_t$ particles yield which leads to an overestimation of the mean transverse momentum $<p_t>$ and of mean $<p_t>$ as a function of charged multiplicity. In the EPOS (39) model a strong increase of the strange particle ratio is obtained if the production of miniplasma is considered in $pp$ collisions. Recently (29) the same model by hydrodynamic evolution simulated in a fairly good way the CDF data on $\Lambda_0$, $k_0$ and charged particles $<p_t>$ vs n (fig.51). It is not clear if the model predicts a critical point for phase transition and at which charged rapidity density. In fig. 51 there is no clear slope change in the $<p_t>$ vs n dependencies. In our view, it would be very interesting to see EPOS model predictions with a phase transition constraint at dn/d$\eta \approx 6$.

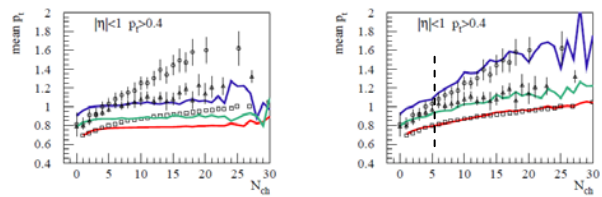

Fig. 51. Average transverse momentum $<p_t>$ as a function of central multiplicity of charged particles for different particle types (from top to bottom: lambdas, kaon short and charged particles). Points are data from CDF experiment. Lines are simulations with EPOS model without hydro (left hand-side) and with hydro (right hand-side). From (29).Dashed line is drawn in correspondence of slope change.

From results of CDF (35), E735 (32), STAR (37) and ALICE experiments, one gets that in the intermediate $p_t$ region not only the $\Lambda^0/K^0$ ratio increases, but the $\Lambda_0$/all particles, the k /π and the $(\bar{p}+p)/(\pi^-+\pi^-)$ increase as well(not shown here). CDF looked at the strange $V_0$ ($\Lambda^0 + K^0$) multiplicity as a function of $< p_t >_{ev}$, the average charged $p_t$ of the event. Fig. 52 shows the average number of $V_0$ per event divided by the average number of $V_0$ in the sample as a function of the $< p_t >_{ev}$. The normalization helps in minimizing $V_0$ finding inefficiencies and make the distributions scale with the energy (40). The maximum value of $\langle nV_0 / \langle n V_0 \rangle \rangle$ is for $< p_t >_{ev} \cong 1$ GeV/c both at 1800 and at 630 GeV.

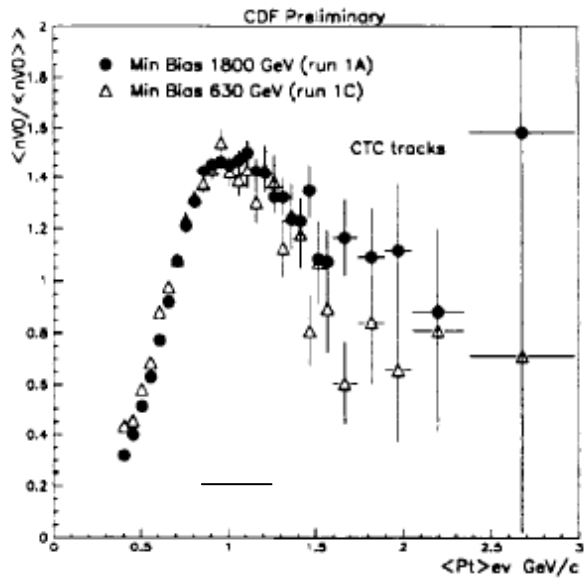

Fig. 52 Profile plot of numbers of $V_0$ versus the mean event $p_t$ (40).

The $< p_t >_{ev} \cong 1$ GeV/c region in CDF data corresponds to events with x=$\langle n/\bar{n} \rangle$ around 1.6, as one can see in fig.53, where n is the charged multiplicity, $\bar{n}$ is the overall mean of the multiplicity distribution and $\langle \rangle$ brackets mean average on events laying in each slice of 0.1 GeV/c size of the $< p_t >_{ev}$. The region with x = 1.6 corresponds to multiplicity n between

5 and 9 in the $< p_t >$ vs n plot (40).

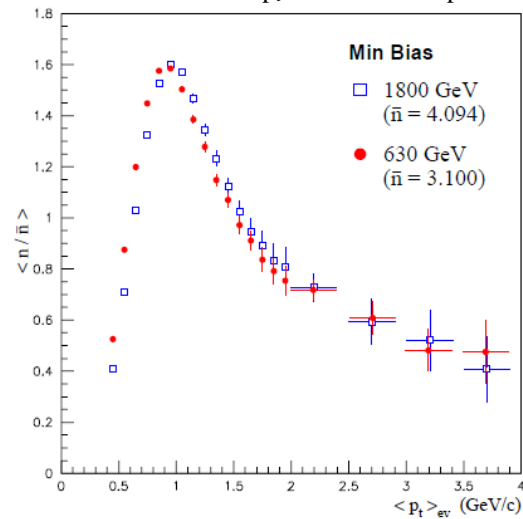

Fig. 53. Profile plot of multiplicity versus the mean event $p_t$ (40).

To conclude: from CDF results we get that at n≅ 6 there is a slope change in the $< p_t >$ vs n dependence for charged particles, for $\Lambda^0$ and for $K^0$, both at 630 and 1800 GeV. For n>6 the $\Lambda^0$ and $K^0$ $< p_t >$ rise only in "hard" events and the increasing is much faster for $\Lambda^0$ $< p_t >$ than for $K^0$. In CDF and in STAR data we see that the "hard" events contain in percentage more kaons and lambdas than "soft" events. The baryon/meson ratio, measured by means of lambda/kaons and antiprotons/pions is higher in "hard" than in "soft" events and is increasing with ρ (21). The maximum "content of strange $V_0$" is in events with $< p_t >_{ev} \cong 1$ GeV/c (not corrected for $p_{t\,min}$ cut) which have multiplicity between 5 and 9. These results suggest that in the events with multiplicity close to the multiplicity of the $< p_t >$ vs n slope change something new happens in the strangeness and baryon production and in the transverse particles motion.

### B. MULTIPLICITY DISTRIBUTIONS.

In the hypothesis that slope change in $< p_t >$ vs ρ signifies a change in the particle production mode, one should see some signals in the multiplicity distribution $P_n$ in correspondence to $\rho^*$ value. For example higher order moments of $P_n$ measure event-to-event multiplicity fluctuations. It is expected for example that in case of phase transition the multiplicity fluctuations would be high. In fig.54, the UA5 multiplicity distribution for -5 < η < 5 at 540 GeV is plotted in the KNO form together with lower energies data (41). The mean multiplicity <n> at UA5 energy in this pseudorapidity range is 29 , from (42). We see that there is KNO violation starting from z in the interval (1.6, 1.8) which corresponds to n ≈ 46–52. The dn/dη distribution at 540 GeV is nearly flat in the range |η| < 3 and has a value 1.5 times smaller for 3< |η| < 5: one can calculate roughly the central pseudorapidity density ρ corresponding to multiplicity

n ≈ 46-52 from the equation $(\rho \cdot 6 + 4 \cdot \rho/1.5)$=46-52, which gives ρ=5.3-6.0

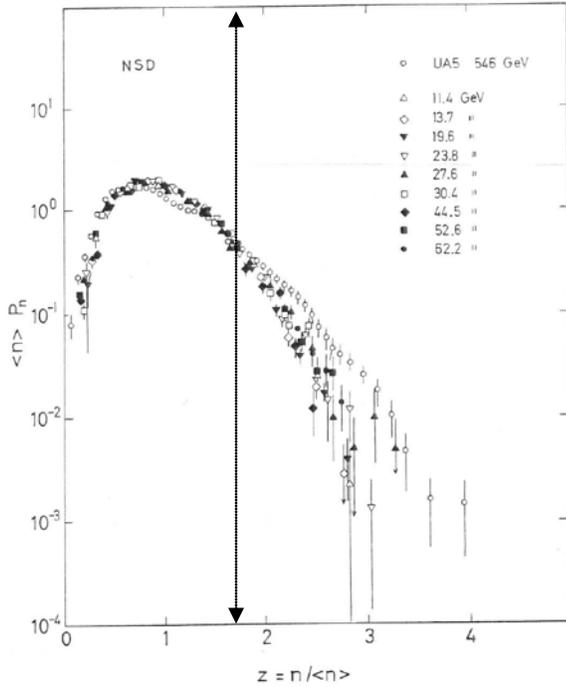

Fig. 54 Charged multiplicities in Non Single Diffractive events plotted as a function of z for UA5 data at $\sqrt{s}$ = 546 GeV, compared to the distributions from the ISR and from Serpukhov and FNAL (41). Dotted arrow points to z value where KNO scaling violation begins.

In fig.55 we present the results of collaboration E735 at 200, 540, 900, 1800 GeV, from (43). The plotted Pn values are for multiplicities n in the full pseudorapidity space. In E735 multiplicity was measured for |η| < 3.25 and extrapolated to the full space by HIJNG Monte Carlo. We remind that the <pt> vs n data of E735 were shown as a function of multiplicity in |η| < 3.25. From dn/dη vs η plots one can deduce that about 64% of the particles are emitted in |η| < 3.25 at 1800 and 900 GeV and about 68% at 540 and at 200 GeV. Pythia Monte Carlo simulations confirmed the 64% percentage at 900 GeV (42). In fig.56 differences between data and the KNO curves in the full space are plotted. We can see that the deviations from the KNO curve begin at the following multiplicities in the full space:

- z = n/<n> = 1.2 at 1800 GeV, with <n> = 44, from which it follows n = 52.8 in full space, which in turn corresponds to ρ ≈ (52.8 x 0.64)/6.50 ≈5.2 in |η| < 3.25.

- z = 1.4 at 900 GeV, <n> = 38, n=53.2,ρ≈5.2;

- z = 1.4 at 540 GeV, <n>=29, n=40, ρ ≈4.2;

- z = 1.7 at 200 GeV,<n>=24, n=41, ρ ≈4.3.

Uncertainty sources on the found ρ values are the extrapolation factor from |η| < 3.25 to full space(± 5%),the multiplicity bin size in KNO plot (± 0 2 5 in ρ) and the systematic on the multiplicity values in the experiment E735,which cancel when we d o comp ae ρ with ρ *.The estimated uncertainty on ρ is about 0.5.

The slope change in the <pt> vs n in the E735 data happens at ρ * = 5.5± 0.7 in |η| < 3.25. In (43), authors put forward the

hypothesis that double parton interactions can account for KNO scaling violation and for the correlation of <pt> with multiplicity n. Here we say something different. We relate KNO scaling violation not to the general <pt> correlation with n, but to the slope change in this correlation. In (43) assumption is made that double parton interactions show up at about $\sqrt{s}$ = 100 GeV, and that at this energy there should be the onset of both KNO scaling violation and <pt> vs n dependence. We point out that at ISR energies there is no KNO violation, at least for most part of the multiplicities, but the <pt> correlates with n (see later).

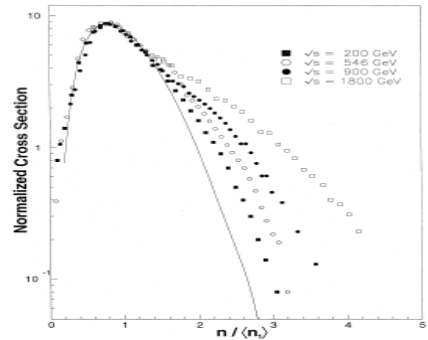

Fig. 55 A comparison of multiplicity distributions at different values of $\sqrt{s}$ from E735. The solid data is the KNO distribution from the ISR data, from (43).

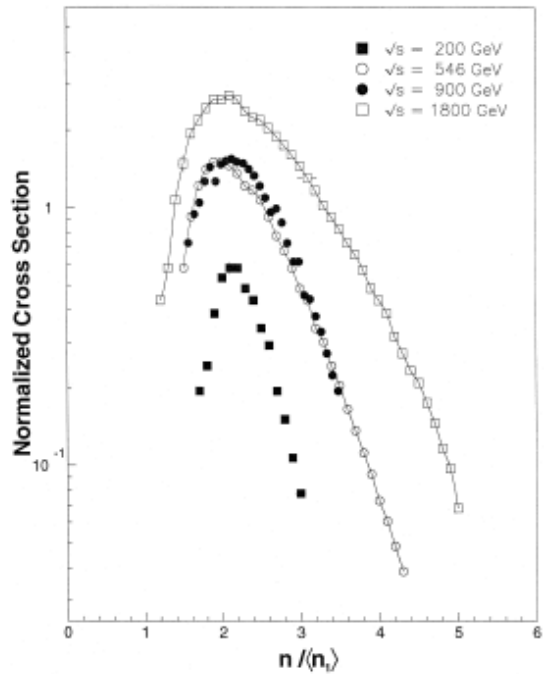

Fig. 56. The derived multiplicity distributions were obtained by taking the difference between the $p\bar{p}$ collider data and the KNO curve.From (43).

Some models (44; 45) explain KNO scaling violation by the fact that the minjets cross section grows with the c.ms. energy. In (42; 45), comparison is made between E735 data at 1800 GeV and a minijets multicomponent model: one can see in fig.57 from residual analysis that there is a strong disagreement between data and model for n ≈ 50, where the model understimates Pn. The multiplicity n=50 at $\sqrt{s}$ =1800 GeV in the full space corresponds to about ρ=5 in $|\eta| < 3.25$,as we saw before.

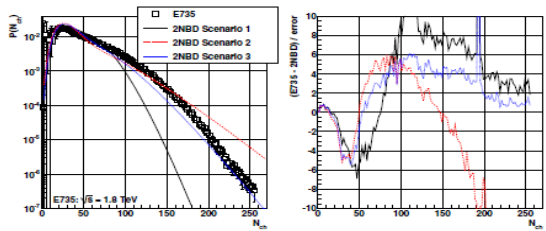

Fig. 57. Comparison between the predictions of the two-component model to the E735 measurement in full phase space at $\sqrt{s} = 1.8$ TeV. The right panel shows normalized residuals between data and the predictions (42).

In the ALICE experiment Pn have been measured at 7, 2.36, 0.9 TeV in $pp$ collisions (27; 46). In fig. 58 and 59 the probability $P(N_{ch})$ of multiplicity $N_{ch}$ in $|\eta| < 1$ is shown. One can see a "slope change" for n between 10 and 15 (ρ between 5 and 7.5), where models at 0.9 and 2.36 TeV, apart from ATLAS-CSC(306), begin to underestimate $P(N_{ch})$. At 7 TeV in these multiplicity interval all shown models underestimate $P(N_{ch})$.

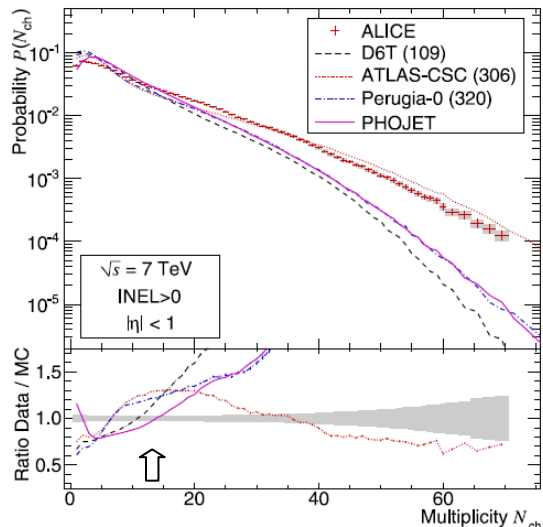

Fig. 58. Comparisons of measured multiplicity distributions for INEL events to models for the pseudorapidity range $|\eta| < 1.0$. The error bars for data points represent statistical uncertainties, the shaded areas represent systematic uncertainties. Data at 7 TeV, from (46). Arrow points to the $N_{ch}$ multiplicity region where most models underestimate the probability $P(N_{ch})$.

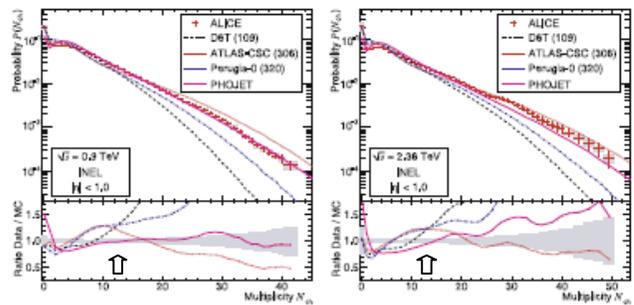

Fig. 59 Comparisons of measured multiplicity distributions for INEL events to models for the pseudorapidity range $|\eta| < 1.0$. The error bars for data points represent statistical uncertainties, the shaded areas represent systematic uncertainties. Left data at 0.9 TeV. Right data at 2.36 TeV (27).The arrows indicate the $N_{ch}$ multiplicity region where most models underestimate the probability $P(N_{ch})$.

In the ATLAS experiment (26) at 900 GeV, one can see (fig.60) how all shown models underestimate the Pn beginning at n=10 , whete the the slope changes in $\langle p_t \rangle$ vs n.

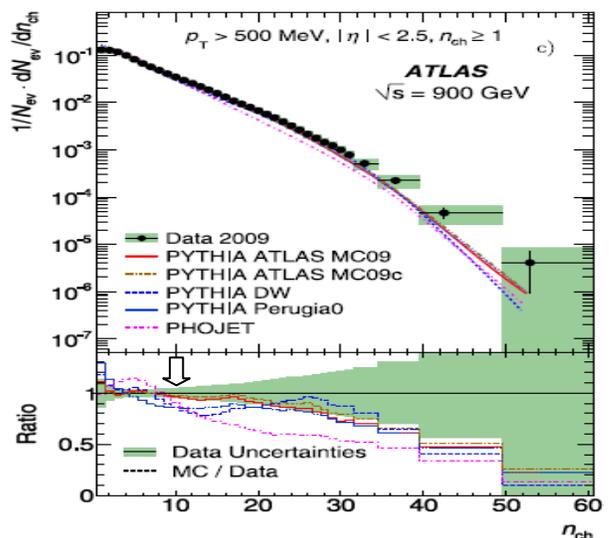

Fig. 60.Charged-particles multiplicities within the kinematic range $p_t > 500$ MeV and $|\eta| < 2.5$. The dots represent the data and the curves the predictions from different Monte Carlo models. The vertical bars represent the statistical uncertainties while the shaded areas show statistical and systematic uncertainties added in quadratures (26).The arrow points to multiplicity value $n_{ch} \approx 10$ where all models begin to underestimate probability of $n_{ch}$.

In the experiment CMS (30) at 7, 2.36 and 0.9 TeV $pp$, the authors note a "slope change" in the Pn distribution for n>20. As shown in fig.61 a for $p_t > 0$,models agree with data up to n≈ 27,where they begin to underestimate the $P_n$. In fig. 61 b, one can see that the same happen at n≈ 10, for $p_t > 0.5$ GeV/c as in ATLAS (the pseudorapidity range is similar in the two experiments, $|\eta|<2.4$ for CMS, $|\eta|<2.5$ for ATLAS).

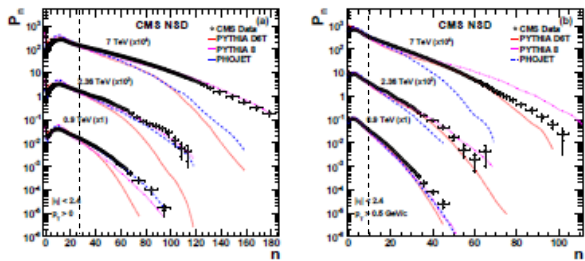

Fig. 61 The charged hadron multiplicities distributions with $|\eta|<2.4$ at $\sqrt{s}=0.9$, 2.36 and 7 TeV, for $p_t>0$ (left panel) and $p_t>0.5$ GeV/c (right panel).Data are compared to two different PYTHIA models and the PHOJET model. For clarity, results for different centre-of-mass energies are scaled by powers of 10 as given in plots (30).Shaded lines show at the three energies the multiplicity values where most models begin to underestimate $P_n$ compared to the data(n≈ 27 for $p_t$ >0 and n≈10 for $p_t$ >0.5 GeV/c)

At ISR energies the slope change in <$p_t$> vs ρ was at ρ ≈ 5 for $|y|<1.5$, which corresponds to multiplicity n of about 27. (We considered ρ constant = 5 for $|\eta|<2$ and ρ = 5/1.5 in average for $2<|\eta|<3$, which gives n = 5 · 4 + (5/1.5) · 2 = 27)(see fig. 8 of (42) ). Because at ISR the <n> is 11 at 50.3 GeV (42), one should expect a KNO scaling violation beginning about at z=27/11=2.5 upward. From fig .62 one cannot make any conclusion due to the experimental uncertainties.

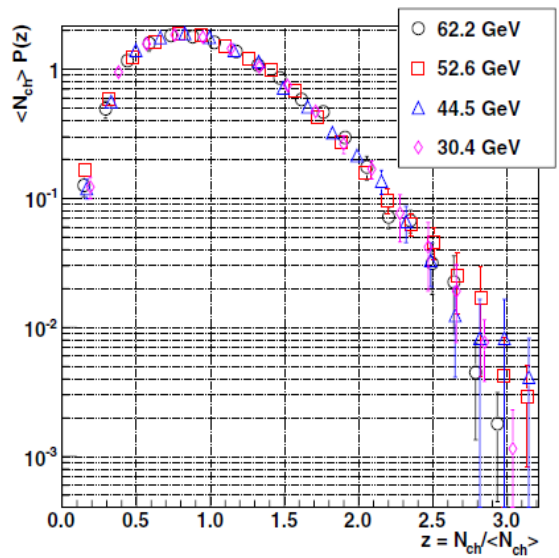

Fig. 62. KNO scaling at ISR energies (42).

In fig.63 $P_n$ from the NA22 at $\sqrt{s}$=22 GeV are shown.
We note that for n=24 and 26, which correspond to ρ > 5, the experimental $P_n$ are much higher than the shown models which fit well data at lower multiplicities.

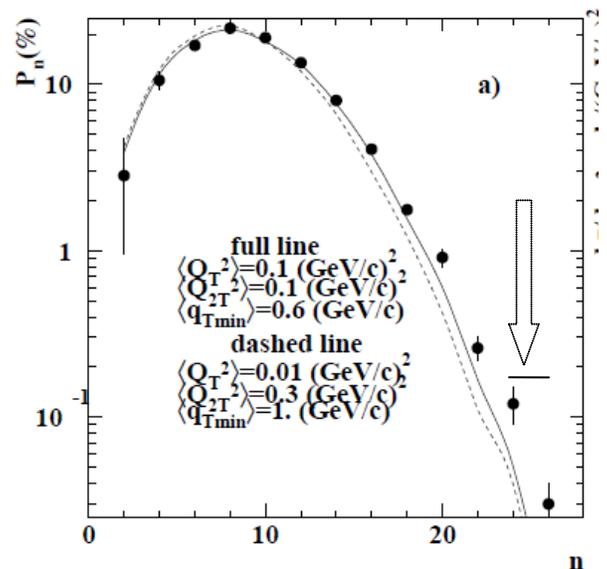

Fig. 63 Multiplicity distributions at $\sqrt{s}=22$ GeV with model comparison. Arrow points to the two points where models deviate from data.

To conclude: it seems that at the multiplicity region where on the <$p_t$> vs n slope change is observed, the experimental $P_n$ becomes higher than both the corresponding KNO values and those predicted by models.This may be due a to new production mechanism, which starts at rapidity density ρ*. This effect may be at least partially responsible for particle overproduction in $pp$ collisions at LHC respect to model prediction:more soft particles are produced compared to models.

Because ρ* is invariant at different c.m.s. energies, one should expect KNO scaling violation to begin at z*=n*/<n>, where n* = ρ*·Δη and Δη is the rapidity interval size. Because <n> grows with c.m.s energy, z*should decrease with c.m.s. energies, and n* will not change with energies.

## C. BOSE EINSTEIN CORRELATIONS, SOURCE DIMENSIONS, COHERENCE OR CHAOTICITY PARAMETER.

The Bose Einstein correlations between like-charged pairs of particles are used to determine the size of the emitting regions (for a review see (47)). Several parametrizations of the ratio between like pairs of charged particles and a reference distributions in terms of the variables constructed from the particle momenta were proposed. Different values of source size are obtained using different parametrizations. In this work we are mainly interested to the dependence of source size on multiplicity and on the $k_t$, the transverse momentum of the particle pairs. Many parameters values can be extracted from data. In this work we consider the following:

- $R_b$ and τ, which are the size and the decay constant of a spherical emitting source;
- the one dimensional invariant radius $R_{inv}$;

- the chaoticity or coherence parameter, which lies between zero (no Bose Einstein effect) and one (maximum chaoticity and maximum interference);

From a parametrization based on Gaussian function one can obtain $R_G$, which is related to $R_b$, that is $R_G \approx R_b/2$

In fig.64 from (47) one can see that in many experiments the source dimension as given by $R_G$ increases with the multiplicity.

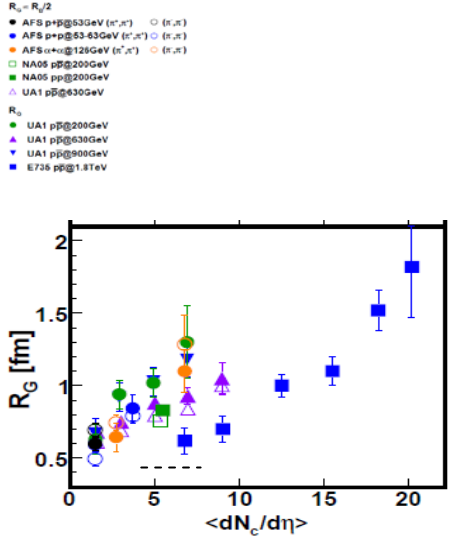

Fig. 64 The multiplicity dependence of $R_G$ radius. Compilation of results from various experiments (47). The dashed line indicates the region of the <$p_t$> vs n slope change.

We are interested at some peculiar behavior in the $\rho$ region where there is slope change in <$p_t$> vs $\rho$, ref. (4). In these region there are only few measures, and we think it will be very interesting to have in near future many experimental points, for example in the $\rho$ region 2-20 with small $\rho$ bin size, preferably within one experiment. It is difficult to put together results from different experiments because of systematic uncertainties due to different ways to extract source dimensions from correlations and to different experimental bias. In fig.65 we plot $R_{inv}$ results from ABCDHW collaboration (7) at 62 GeV $pp$, CMS (48) and ALICE (49) at 900 GEV $pp$, STAR(50) at 200 GeV $pp$, and E735(51) at 1800 GeV $p\bar{p}$. We applied some correction factors to the published data. Namely we multiplied by factor 1.25 the $\rho$ values for ABCDHW collaboration, to take into account the acceptance losses, as we did for the <$p_t$> vs n. We divided by $\sqrt{\pi}$ the R values of CMS as in (49) because of different parametrization.

From fig.64 and fig.65 one can see an overall trend. The source size increases with multiplicity in a nearly linear way up to $dn_{ch}/d\eta \approx 5$, where it may be possible a different data trend. As we sad before, more data are needed to study some possible structures in the rising.

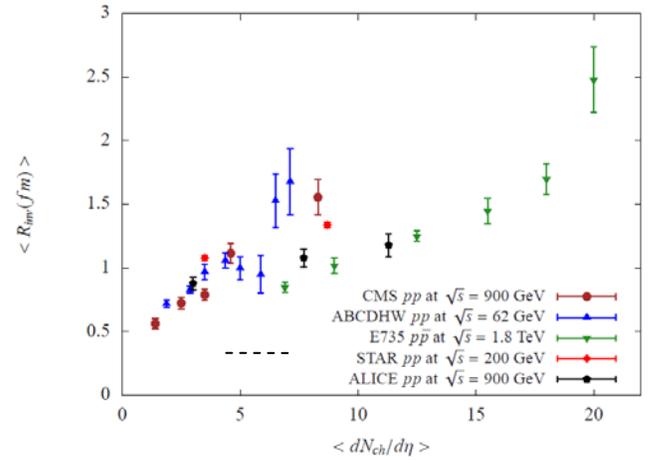

Fig. 65 Multiplicity dependence of Rinv from various experiments. The dashed line indicates the region of the <$p_t$> vs n slope change.

The chaoticity parameter decreases with $\rho$, as one can see from results of UA1(52), E735, ABCDHW, CMS collaborations, not shown here.

E735 ,STAR and ALICE collaborations studied the dependence of radii on the transverse momentum $k_t$ of the particle pairs. At lower energy, a similar study has been done by NA22 and NA27 experiments (47). In (49), ALICE collaboration shows an analysis of results. The r correlation with $k_t$ depends strongly on the used background reference distribution. In fig.66 from (49) we can see that both STAR and E735 find that $R_{inv}$ decreases when $k_t$ increases. ALICE data does not show this trend.

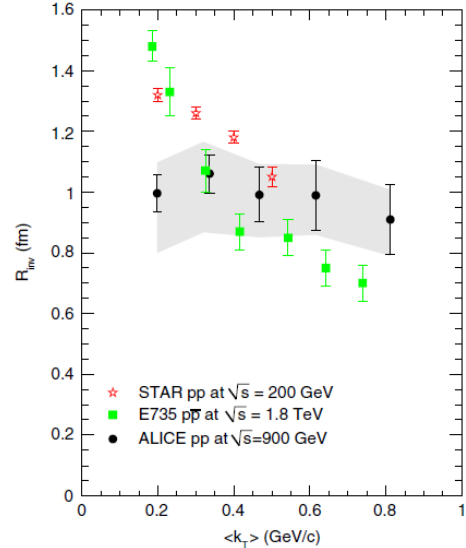

Fig. 66. One-dimensional Gaussian HBT radius in $pp$ collisions at $\sqrt{s} = 900$ GeV as a function of transverse momentum $k_t$ from ALICE collaboration (full dots). Stars and filled boxes represent the radii measured by STAR coll. and E735 coll., respectively (49).

We suggest a possible interpretation. ALICE data in fig.66 are from a sample with <p> = 3.6. E735 events have <p> = 14.4 . Considering ref. (50) from STAR experiment one can deduce that since particle pairs enter into the correlation function, the effective average multiplicity at midrapidity is about 6.0 in $|\eta| < 0.5$. In fig.67 from ALICE one can see that $R_{inv}$ dependences on $k_t$ for multiplicity $\leq 6$ ($\rho \leq 3.7$) and $\geq 7 (\rho \geq 4.4)$ are different. In higher multiplicity events there is a decreasing of $R_{inv}$ vs $k_t$. There may be a physical interpretation to these results. At low impact parameter there is an higher strings overlap, which gives higher particle density $\rho$, higher $R_{inv}$ as a measure of the overlapping region and an increase of $<p_t>$ (1). If a critical overlapping is reached it may happen that for some $\rho$ values, $R_{inv}$ does not increase because some kind of "maximum" overlap has been reached. Higher $\rho$ values may correspond to quark gluon plasma production in an expanding volume. This volume may be increasing with the multiplicity. For $\rho$ values below the critical region there may be weak or no dependence of $R_{inv}$ on $k_t$ because all particles are emitted from the same interaction region. In events with $\rho$ higher than the transition value, fireball expansion correlates high energy particles with earlier times when fireball was both smaller and at a higher temperature. One expects that the difference between $R_{inv}$ at small $k_t$ and at high $k_t$ should be higher at higher $\rho$, due to longer expansion. As one can see from fig.66 at low $k_t$ the radii measured by STAR and by E735 are about 1.5 times higher than the radius measured by ALICE at the same $k_t$, consistent with the fact that the final particles source increases with $\rho$.

particles. We probably saw already this effect on E735, CDF and UA1 data.

## D. *Event by event fluctuation in the average event $p_t <p_t>_{ev}$.*

At CDF (RUN I) at 1800 and 630 GeV an interesting study has been done in minimum bias events (14). The studied quantity was the dispersion $D_m$ of the mean event $p_t$ for events with multiplicity m. $D_m$ is defined by

$$D_m(\bar{p}_t) = \frac{\langle \bar{p}_t^2 \rangle_m - \langle \bar{p}_t \rangle_m^2}{\langle \bar{p}_t \rangle_{sample}^2}$$

where brackets $<>$ indicate average over all events with a given multiplicity m, $\bar{p}_t$ is the mean event $p_t$. In fig .68 $D_m$ is plotted as a function of the inverse of the multiplicity. The dispersion $D_m$ is expected to decrease with increasing multiplicity and to converge to zero when m goes to infinity if only statistical fluctuations are present. An extrapolation to a non zero value would indicate non statistical fluctuation from event to event in the $p_t$ distribution. As pointed out in (14), the points deviate from linearity at multiplicity $\geq 7$. Data show that there are non statistical fluctuations and seem to indicate the onset of a distinct type of event by event fluctuations at multiplicity $\geq 7$.

The deviations from the linearity are more visible in hard events(fig.70) than in the soft events(fig.69).

At both energies one can see for $n \geq 7$ a drop of points from linearity. At 1800 GeV, where statistics at high multiplicity is better, it is possible to see a kind of plateau for events with n from 9 up to 14.

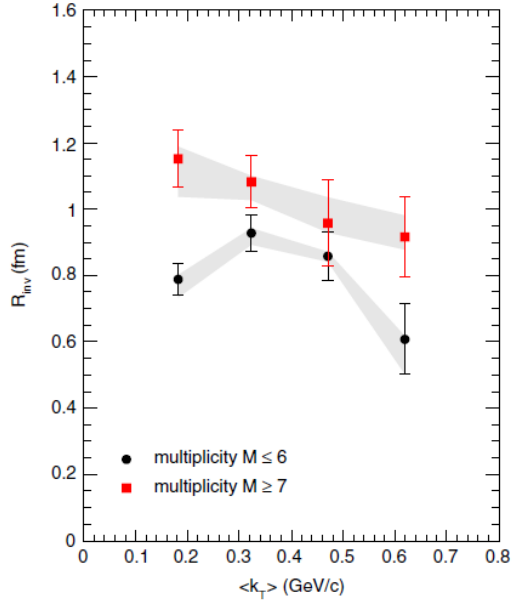

Fig. 67. One dimensional radius as a function of $k_t$ for low (black circles) and high (red squares) multiplicity events, from ALICE collaboration (49).

The slope change in the $<p_t>$ vs $\rho$ for $\rho$ higher than $\rho^*$ may be due to the fact that the particle emission is different in the different $\rho$ regions and that hydrodynamical effects show up, which may be seen in $<p_t>$ vs $\rho$ dependence for identified

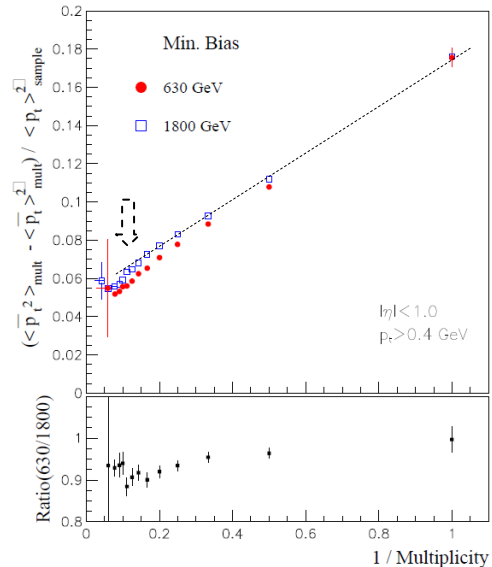

Fig. 68.Dispersion of the mean event $p_t$ as a function of the inverse multiplicity for the full minimum bias samples at 1800 and 630 GeV by CDF collaboration. At the bottom the ratio of the two curves is shown (14).Shaded arrow points to the region where data drop from linearity.Shaded line is for reader eye guiding for the 1800 GeV data.

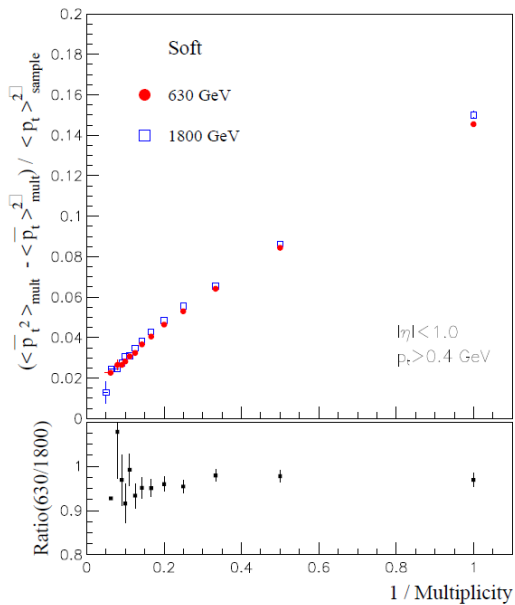

Fig. 69. Dispersion of the mean event $p_t$ as a function of the inverse multiplicity for soft events samples at 1800 and 630 GeV by CDF collaboration. At the bottom the ratio of the two curves is shown (14).

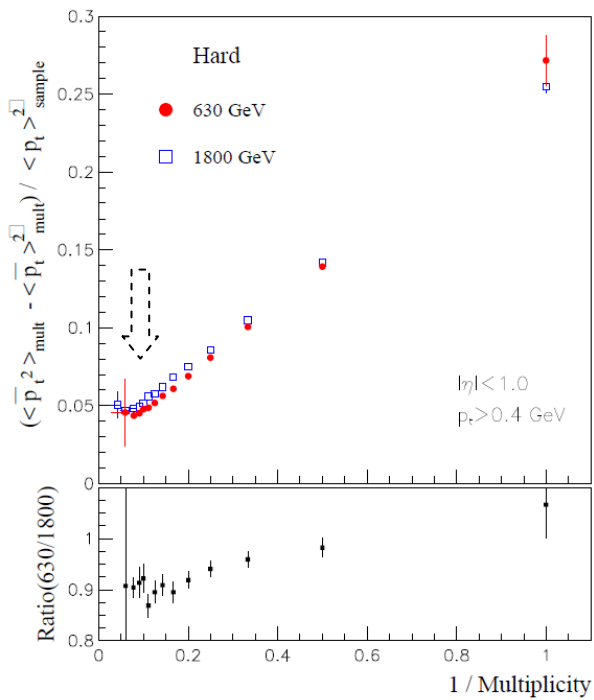

Fig. 70 Dispersion of the mean event $p_t$ as a function of the inverse multiplicity for the hard events samples at 1800 and 630 GeV by CDF collaboration. At the bottom the ratio of the two curves is shown (14). Shaded arrow points to the region where data drop from linearity.

In fig .71 the dispersion $D_m$ vs the inverse of multiplicity n is plotted for data from ISR experiment in $pp$ collisions at 62 GeV (53). One can see behavior similar to CDF study: for multiplicity n higher than 20, points deviate strongly from linearity. There is a plateau of 7 points from n=20 up to n=26.

The ISR measurement has been done in the region $|x| = |p_{||}| / |p|_{max}| < 0.3$, which corresponds to about $|\eta| < 3.6$ at $\sqrt{s} = 62$ GeV. The acceptance correction factor, which was not applied to data, is about 1.4, that means that for the corrected n values the plateau begins at n = 37. The pseudorapidity density at 63 GeV has roughly a constant value for $-2 < \eta < 2$ and for $2 < |\eta| < 3.6$ has an average value of one half the constant central value $\rho$. It results that multiplicity n=37 corresponds to $\rho$=6.6. To conclude: it seems that there is an onset of a distinct type of event by event fluctuation of $\langle p_t \rangle_{ev}$ in events with $\rho > \rho^*$ in comparison to lower multiplicity events.

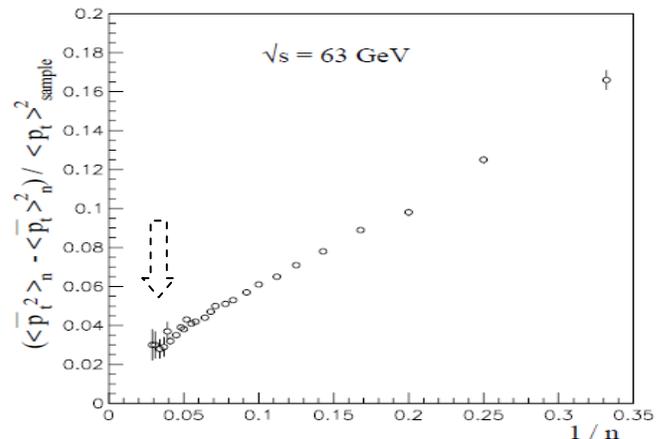

Fig. 71 The mean event $p_t$ dispersion at 63 GeV c.m.s. energy (as measured at ISR by SFM) is plotted against the inverse multiplicity 1/n (53). Shaded arrow points to the region where data deviate from linearity

### E. LONG RANGE RAPIDITY CORRELATIONS.

In the study of long (pseudo)rapidity correlations the average number $\langle n_b \rangle$ of charged particles in the backward (pseudo)rapidity region is plotted against the number $n_f$ of charged particles in the forward( pseudo)rapidity region. In fig. 72 we present UA5 results at $\sqrt{s} = 540$ GeV (41). In this plot forward region is $0 < \eta < 4$, backward $-4 < \eta < 0$. There is a linear relation between $\langle n_b \rangle$ and $n_f$, but one can note that a $n_f$= 25, where $\langle n_b \rangle \approx 20$ there is a shoulder and the relation deviates strongly from linearity, for about 7 points. The average $n_f + \langle n_b \rangle = 45$ corresponds to $dn/d\eta = 45/8 = 5.6$. If we take into account that at 540 GeV $\rho$ is about constant for $-3 < \eta < 3$ (42) with value $\rho$ and in interval $3 < |\eta| < 4$ has a value $\rho$=0.8 $\rho$, we can estimate the central $\rho$ corresponding to 45: $(6 \cdot \rho + 0.8 \cdot \rho \cdot 2) = 45$ which gives $\rho = 5.9$.

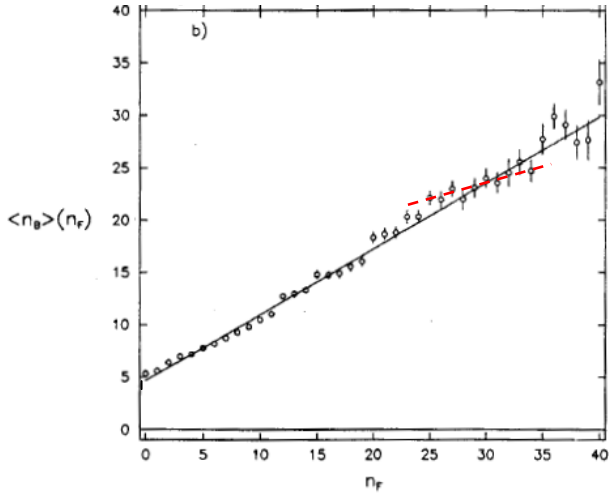

Fig. 72.The relation between the average of $n_b$ (at fixed $n_f$) and $n_f$. Region f : $0 \leq \eta \leq 4$ ,Region b : $-4 \leq \eta \leq 0$. UA5 collaboration (41).Shaded line shows a possible deviation from a linearity

An interesting quantity to study is $\langle z^2 \rangle$, with $z = n_f - n_b$.

This quantity has been measured among others by the E735 collaboration at 1800 GeV (54). As pointed out by authors and in (55), there is a slope change or an hump for $n \cong 40$, which corresponds to $dn/d\eta = 6.2$. In (54) the fit to the data has been done with cluster model: for multiplicity $n = 40$ and higher is necessary to introduce a cluster with size 4.0, which replaces the cluster of size about 2.6 with which they fit data up to $n = 40$ (fig .73). In fig.74, data from E735 at different energies together with linear model fit are shown.

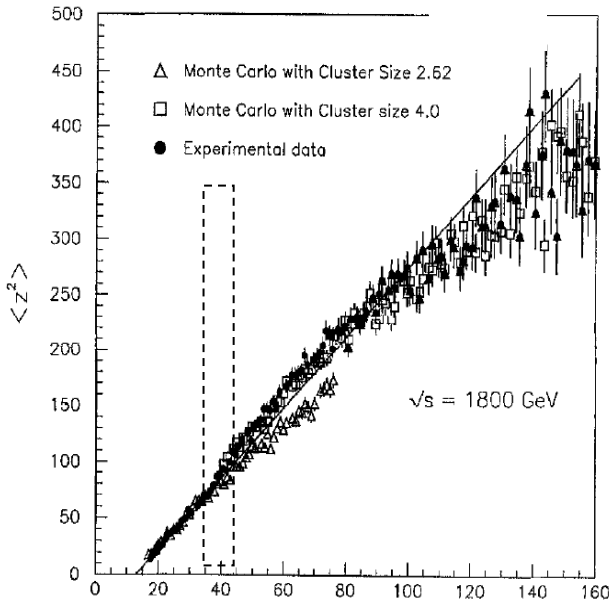

Fig. 73. Variance of the asymmetry distribution for $|\eta| < 3.25$ at 1.8 TeV as a function of the total charged multiplicity (54).Dashed box shows the "hump" region.

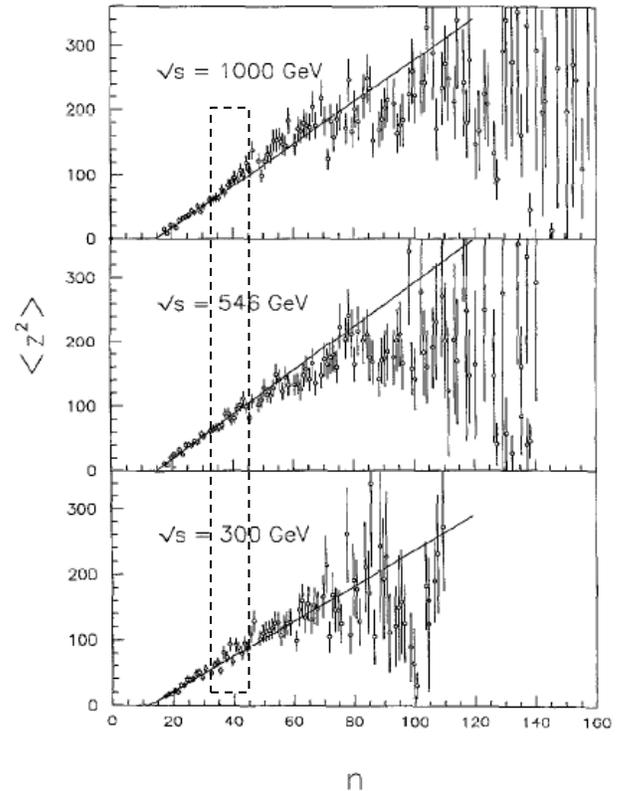

Fig. 74 Variance of the asymmetry distribution as a function of the total charged multiplicity at the listed energy for $|\eta| < 3.25$. The solid line is a fit up to $n \approx 40$. E735 coll. (54).

It seems that at $\rho$ values where there is slope change in $\langle p_t \rangle$ vs n there are also changes in the behavior of forward-backward rapidity correlations. A possible explanation is the following: the $\rho$ increasing is due to many strings on top of each other and the clustering of strings may eventually lead to the formation of a macroscopic cluster at a certain critical density: this would indicate the onset of a phase transition. When strings fuse, a reduction in the long range correlation may be expected.

The CMS collaboration (56) has recently released the first study of angular correlations from $pp$ collisions at LHC. A long range, near side correlation increases in strength with increasing multiplicity and is most prominent in the region $1 < p_t < 3$ GeV/c. The effect is more visible at high multiplicity. A multiplicity of 35 in the phase space $|\eta| < 2.4$ $p_t < 0.4$ GeV/c corresponds to $\rho \cong 14$. At lower multiplicity the effect is weaker but it seems there is still a significant difference of data from models (fig.75). It would be interesting to determine at which value of $dn/d\eta$ a weak but significant shifting from the models shows up. The effect is strong and significant at intermediate $p_t$ (1-2 GeV/c) where, as we said before, there is high k/pion, lambda/kaon and antiproton/pion ratio. It is prominent in high multiplicity events, where presumably the source radii are high, and it may be due to the fact that there is an hydrodinamical expansion as in heavy ions collisions. Further measurements are needed and it would be very interesting to correlate them with source dimension.

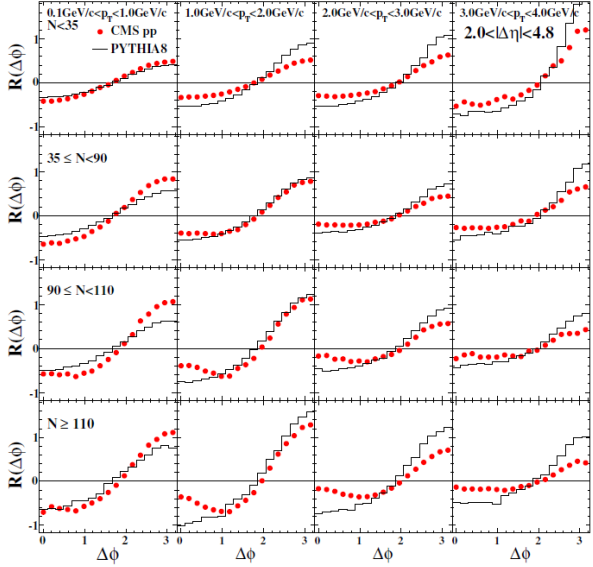

Fig. 75. Projections of 2-D correlation functions onto $\Delta\varphi$ for $2.0<|\eta|<4.8$ in different $p_t$ and multiplicity bins for fully corrected 7 TeV data and reconstructed PYTHIA8 simulations (56).

### F. SHORT RANGE RAPIDITY CORRELATIONS.

In (57), many particles rapidity correlations have been investigated in $pp$ and light ion interactions: the study has been done in different multiplicity bins, and only for the largest bin, with multiplicity > 16, there is a tendency towards small cluster. The rest of bins show a behavior similar to that of the inclusive data. The results are similar in $pp$, $p\alpha$, $\alpha\alpha$. The study has been done for $|y| < 1.5$. The measured multiplicity bin > 16 corresponds to a measured $dn/dy > 5.3$, and taking into account the acceptance correction of 1.4 we get $dn/dy > 7$. In the PHOBOS experiment (58) at RHIC the short rapidity correlation has been measured at c.m.s. energies of 200 and 410 GeV. One can see from fig.76 that as function of $z = n/\langle n \rangle$ the effective cluster size $K_{eff}$ increases up to the points $z = 1.5, 1.8, 2.1$ which corresponds to $\rho$ about 4.5, 5.4 and 6 respectively. In fig.77, data from UA5 at 200, 546 and 900 GeV, SFM at 44 GeV and PSB at 63 GeV are shown.

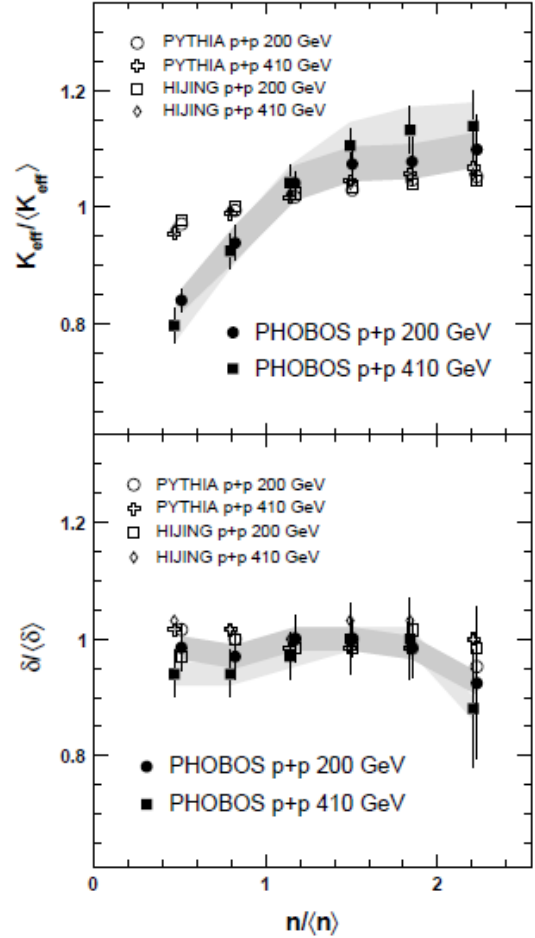

Fig. 76 Normalized effective cluster size $K_{eff}/\langle K_{eff} \rangle$ (top) and decay width $\delta/\langle\delta\rangle$ (bottom) as a function of normalized multiplicity $n/\langle n \rangle$ in $p+p$ collisions at $\sqrt{s} = 200$ and 410 GeV, measured by PHOBOS (solid symbols), as well as MC studies (open symbols). PHOBOS coll. (58).

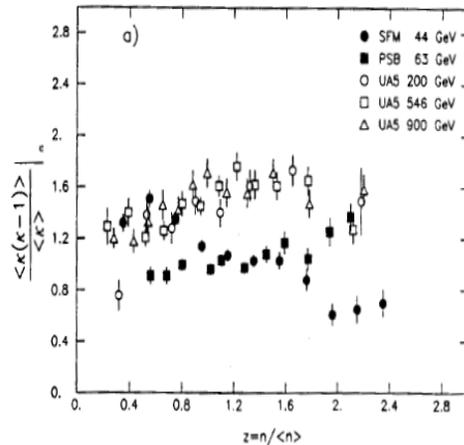

Fig. 77 The effective $K_{eff}$ is plotted versus the scaling variable $z = n/\langle n \rangle$ for the ISR energies 44 and 63 GeV and for 200, 546 and 900 GeV.

The $K_{eff}$ trend versus the multiplicity in UA5 data is similar to the one in heavy ion collisions, where $K_{eff}$ is plotted versus the event centrality (59).

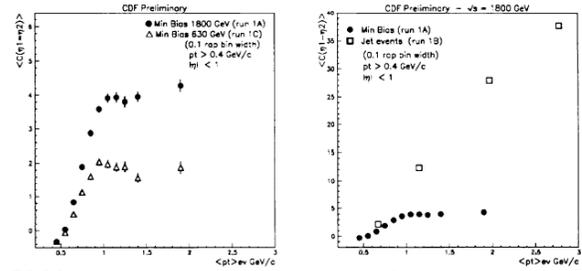

Fig. 79 Mean strength of the correlation is plotted as a function of the $<p_t>_{ev}$. Panel(a): minimum bias at $\sqrt{s} = 1800$ GeV (black points) and at 630 GeV(open triangles). Panel(b) As in panel (a) for 1800 GeV minimum bias (black points) and 1800 GeV jet events, with at least one jet with energy > 15 GeV (60).

In conclusion, it seems that there are changes in short range rapidity correlation strength for $\rho \approx \rho^*$.

## G. SPIKE EVENTS.

From experiment NA22 (61) at $\sqrt{s} = 22$ GeV $\pi^+ p$ interactions we report fig.80, where the average values of the maximum density in pseudorapidity in $d\eta = 0.5$ is plotted at each multiplicity n. One sees that the point at n = 24 deviates substantially from the linear trend of the lower multiplicity data. We saw already in the paragraph on multiplicity distributions that Pn for n = 24 and n = 26 were in strong disagreement with a fit to lower multiplicities data. The multiplicity n = 24 corresponds to $dn/d\eta \cong 5$, because at 22 GeV $dn/d\eta$ is about flat for $\eta$ from -2 to +2 and goes down to 0 at about $\eta = 4$.

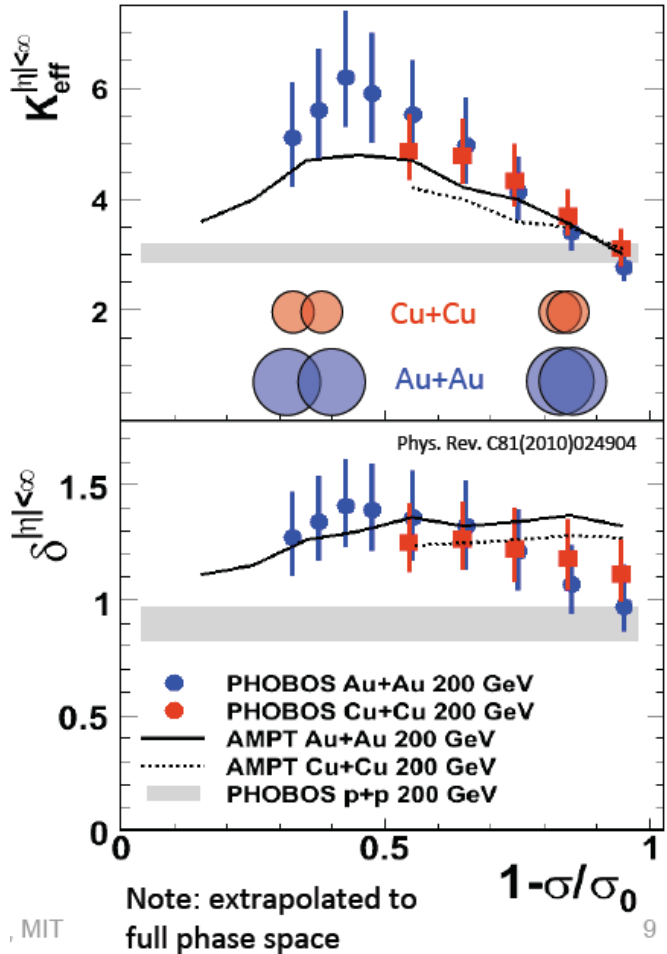

Fig. 78.Cluster dimension $K_{eff}$ (upper panel) and cluster width as a function of fractional cross section for PHOBOS data and for the AMPT model in Cu+Cu and Au+Au collisions for $|\eta| < 3$ at $\sqrt{sNN} = 200$ GeV (59).

The short range rapidity correlation as function of $<p_t>_{ev}$ has been studied by CDF at 630 and 1800 GeV (60): for $<p_t>_{ev}$ around 1 GeV/c the correlation strength C stops growing (fig. 79). In fig (79 a), C values vs $<p_t>_{ev}$ are plotted for 630 and 1800 GeV minimum bias events. In fig (79 b) results are shown again for the 1800 GeV minimum bias and for events from a high ET jet trigger: we see that the C correlation strength increases steadily with $<p_t>_{ev}$ in jet trigger events. The slowing down in minimum bias of the correlation strength happens at $<p_t>_{ev} \cong 1$ GeV/c, which corresponds to events with multiplicity range 5-9.

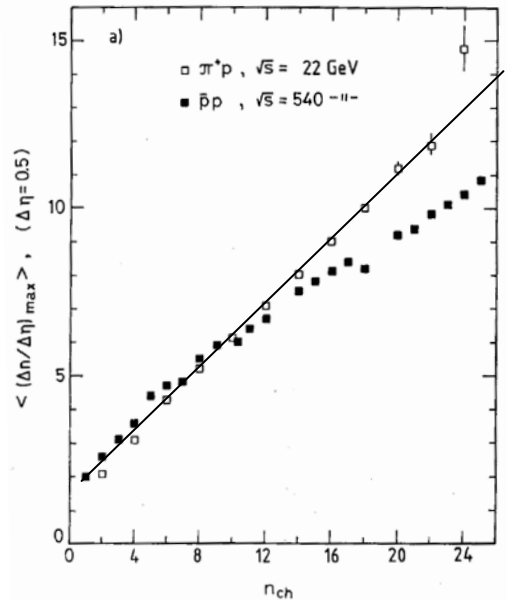

Fig. 80. The maximum track density$(dn/d\eta)_{max}$ within an event, averaged over all events with given charged multiplicity $n_{ch}$, as a function of $n_{ch}$.The $\pi + p$ data at $\sqrt{s} = 22$ GeV are compared to $pp$ data at 540 GeV, the pseudorapidity window is $d\eta = 0.5$ (61).

In fig.81 one sees one NA22 event whose probability to happen for statistical fluctuation has been calculated to be very low. The event has n = 26 charged particles. From the picture one can deduce that dn/dy averaged on the full rapidity interval is 6.5. In an interval of dy = 0.1 there are 10 particles, which correspond to dn/dy = 100, and these 10 particles are uniformly distributed in azimuth φ. In azimuth we can see that there are two narrow bands in φ of about π/4 each of which extends for about dy = 3. This event has a strong particle rapidity correlation and a strong particle azimuthal correlation which extends for a large interval in y.

*H. CORRELATIONS BETWEEN THE PRODUCTION OF PROMPT POSITRONS AT LOW TRANSVERSE MOMENTUM AND THE ASSOCIATED CHARGED MULTIPLICITY.*

At ISR the $e^+/\pi$ ratio has been studied for identified positrons for different slices of $p_t$. From fig.82 one sees that for positron with $p_t < 0.4$ GeV/c the ratio increases linearly as a function of total energy $E_{tot}$ measured in a calorimeter. The results are quoted for different $E_{tot}$ intervals, with the corresponding <n> multiplicities. The $e^+/\pi$ ratio increases linearly with <n>, which means that the positrons number increases quadratically with multiplicity. This may be a Quark Gluon Plasma signal or soft −annihiliation of quarks and antiquarks. (62).

We can see from fig.82 that there is no increase of the ratio for $p_t > 0.4$ GeV/c. We note that the ratios for $p_t < 0.4$ GeV/c and that for $p_t > 0.4$ GeV/c are nearly equal for $2 < E_{tot} < 5$ GeV, and they are different in the last three $E_{tot}$ bins, which correspond to average multiplicities <n> of 6.4, 9.4 and 11.6, respectively. The rapidity interval was |y|<1. One can see in fig.83 that for $E_{tot} < 5$, all events have multiplicity < 12 (ρ<6). An hypothesis that may be tested is that the positron/pion ratio at low $p_t$ has an high value only in events with $\rho > \rho^*$. At the time of writing experimental results are very few and no conclusion can be drawn.

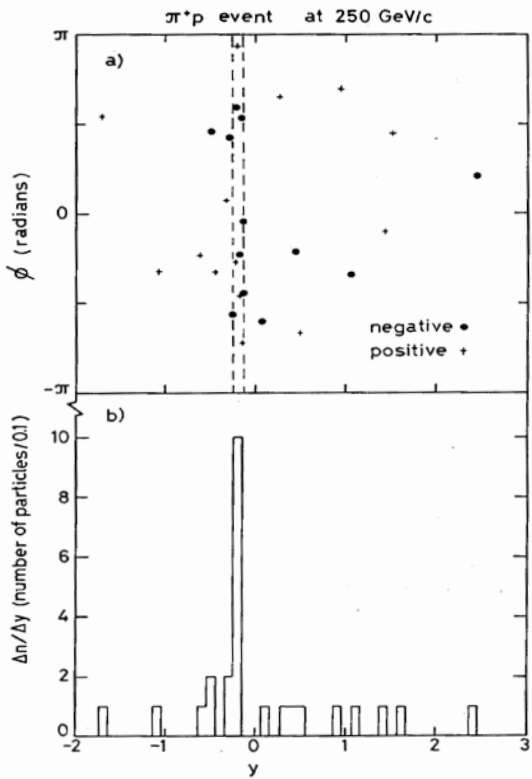

Fig. 81,Scatter plot of azimuthal angle φ versus c.m. rapidity for all secondary charged particles in the anomalous π event (61).

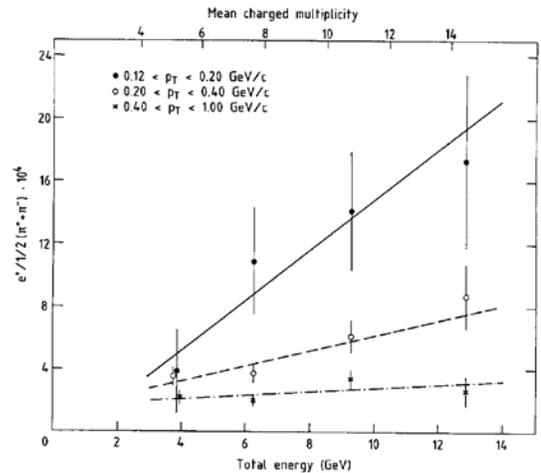

Fig. 82 $e^+/\pi$ ratio as a function of total energy, measured with calorimeter, for 3 $p_t$ intervals. The top scale indicates the mean charged multiplicity which corresponds to the measured Etot. AFS coll. (62).

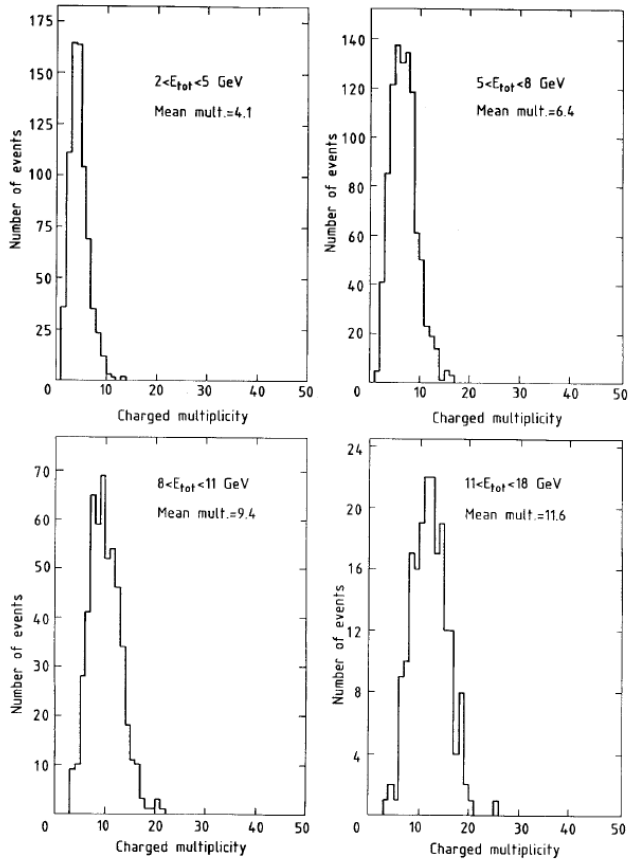

Fig. 83 Multiplicity distributions for the four different $E_{tot}$ bins (62).

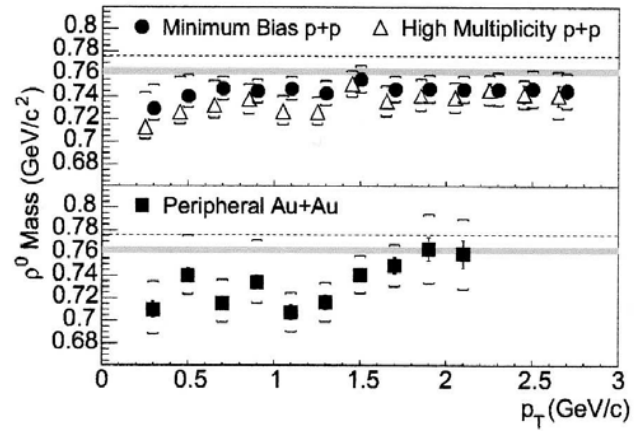

Fig. 84. The $\rho^0$ mass as a function of $p_t$ for minimum bias p+p (filled circles), high multiplicity $p+p$(open triangles), and peripheral Au+Au(filled squares) collisions at $\sqrt{sNN} = 200$ GeV. The dashed lines represent the average of the $\rho^0$ mass measured in $e^+e^-$. The shaded areas indicate the $\rho^0$ mass measured in $p+p$ collisions at $\sqrt{s} = 27.5$ GeV (63).

## I. $\rho^0(770)$ MESON MASS SHIFT.

The STAR experiment in $pp$ collisions at 200 GeV (63) has measured the $\rho^0(770)$ resonance production. From fig. 84 we can see that in minimum bias $pp$ events there is a shift in the measured mass value of about 20 MeV/c$^2$ compared to the expected value. A more consistent shift of about 50 MeV/c$^2$ is seen in high multiplicity $pp$ events. The high multiplicity sample correspond to 10% events with highest multiplicity. As an estimate we can say that the 10% highest multiplicity events has z > 1.7 (41). The <dn/dη> in 200 GeV minimum bias events is about 3.2.This means that the high multiplicity sample contains events with dn/dη >(1.7 x 3.2)= 5.4. In the high multiplicity sample, the $\rho^0$ mass shift is comparable to the mass shift measured in Au-Au collisions at RHIC. Some comments are in order. The mean path of the $\rho^0$ resonance is about 1.2 fm. In high multiplicity $pp$ we saw that the source size is higher than 2 fm. A possible explanation is that the $\rho^0$ resonance mass shift is due to $\pi$ rescattering in the medium (64). It would be interesting to measure the mass shift for many values of $dn/d\eta$.

## IV. CONCLUSIONS.

The analysis of correlation of <$p_t$> with charged pseudorapidity central density $\rho$ shows that at energies from 22 GeV to 7 TeV in different experiments, there is a slope change at $\rho$ value $\rho^*$, which is constant within systematic errors. The slope change is seen in 21 $p_t$ vs n curves, in different experiments and mostly is not predicted by models. The overall experiments average value is $\rho^* = 5.5$, with a standard deviation of 0.6. Many interesting phenomena seem to depend on the ratio R = $\rho/\rho^*$. The <$p_t$> vs $\rho$ dependence for identified particles shows slope change at R=1. For R>1 is more evident the dependence on particle mass of the increasing rate of <$p_t$> vs $\rho$. In events with R ≥ 1 the k/$\pi$, $\Lambda$/k and $\bar{p}/\pi$ ratio is higher than in events with R < 1. For R ≥ 1 the multiplicity distributions at many energies begin to show KNO scaling violation and deviation from most model prediction. The source dimension r as measured by Bose Einstein correlations is increasing with $\rho$ and there are hints from data that the increase is different for R < 1 ad R > 1. The dependence of r on particle pairs transverse momentum seems to be different for events, depending on R. The <$p_t$>$_{ev}$ fluctuations deviate strongly from the linearity for R ≥ 1. The forward backward correlation shows a slope change and the variance of the asymmetry dispersion($z^2$) shows an hump for R ≈ 1. The strength of short range correlations stops to grow near R = 1 and at <$p_t$>$_{ev}$ value found mostly in events with R ≥ 1. There are few hints that high rapidity density fluctuations, low $p_t$ electron production and $\rho^0$ (770) resonance mass shift may depend on R.

Models which predict an increasing with multiplicity of the percentage of perturbative QCD minijets explain well many of these phenomena but fail to describe sudden changes, like the $<p_t>$ vs n slope change. These models predict a gradual transition from soft events to events with jet structure. In our opinion data don't show that at least at intermediate $p_t$ the jet structure is dominating. It is conceivable that high percentage of events with medium-low transverse energy $E_t$ are still "soft" (non perturbative QCD) events, with possibly a new particle production mechanism.

A simple description may be the following. As the impact parameter decreases ,strings overlap increases. This may cause an increase of interaction radius, average transverse momentum and multiplicity. If a critical overlap is reached, a phase transition may happen, with a change in the $<p_t>$ vs n slope due to a change in the particle production and emission mechanism. At transition, strong event by event fluctuations may happen, both in $< p_t >_{ev}$ and multiplicity. For $\rho$ values in the transition region the source radius r may increase weakly or even stay constant. For higher $\rho$, r may increase in a new way due to plasma expansion, with a dependence on the transverse momentum $k_t$ of particle pairs. Hydrodynamical effects may cause change in particle correlations and on the dependence on particle mass of the correlation of $<p_t>$ with multiplicity. In a dense medium the production probability of baryons ,strange particles and low $p_t$ positron may increase and particle rescattering may be more likely,with a possible effect on short resonance mass value.

## V. ACKNOWLEDGEMENTS.

We thank dr.N. Moggi and dr.G. Ferri for helpful comments and data supply and mr. Tiziano "Drakulone" Serafini for enlightening discussions.